\PassOptionsToPackage{table}{xcolor}
\documentclass[10pt,twocolumn,letterpaper]{article}

\usepackage{cvpr}              
\usepackage{multirow}
\usepackage{rotating}
\usepackage{tabularx}
\usepackage{multicol}
\usepackage{threeparttable}

\newcommand{\change}[1]{\textcolor{black}{#1}}
\usepackage[dvipsnames]{xcolor}

\definecolor{cvprblue}{rgb}{0.21,0.49,0.74}
\usepackage[pagebackref,breaklinks,colorlinks,citecolor=cvprblue]{hyperref}

\def\paperID{15} 
\def\confName{CVPR}
\def\confYear{2025}

\title{VolTex: Food Volume Estimation using Text-Guided Segmentation and Neural Surface Reconstruction}

\author{Ahmad AlMughrabi\\
Universitat de Barcelona, Spain\\
{\tt\small ahmad.almughrabi@ub.edu}
\and 
Umair Haroon \\
Universitat de Barcelona, Spain\\
{\tt\small umairharoon@ub.edu}
\and
Ricardo Marques$^*$\\
Grup de Tecnologies Interactives (GTI),\\ 
Universitat Pompeu Fabra (UPF), Spain\\
{\tt\small ricardo.marques@upf.edu }
\and 
Petia Radeva$^*$\\
Universitat de Barcelona, Spain\\
IMUB \& Institut de Neurociències, Barcelona\\
{\tt\small petia.ivanova@ub.edu}
}

\begin{document}
\maketitle

\def\thefootnote{*}\footnotetext{Equal supervison.}

\def\thefootnote{\arabic{footnote}}

\begin{abstract}
Accurate food volume estimation is crucial for dietary monitoring, medical nutrition management, and food intake analysis. Existing 3D Food Volume estimation methods accurately compute the food volume but lack for food portions selection. We present VolTex, a framework that improves \change{the food object selection} in food volume estimation. Allowing users to specify a target food item via text input to be segmented, our method enables the precise selection of specific food objects in real-world scenes. The segmented object is then reconstructed using the Neural Surface Reconstruction method to generate high-fidelity 3D meshes for volume computation. Extensive evaluations on the MetaFood3D dataset demonstrate the effectiveness of our approach in isolating and reconstructing food items for accurate volume estimation. The source code is accessible at \footnote{https://github.com/GCVCG/VolTex}.
\end{abstract}    
\section{Introduction}
\label{sec:intro}
Accurate nutrient analysis of food requires determining its weight, which is derived from both its density and volume \cite{liu2020food}. While density estimation can be achieved using standard food density tables, estimating volume from 2D images remains a significant challenge due to the lack of depth information. Since standard mobile phone cameras do not provide depth data, research has focused on deriving 3D volume estimates from \textit{videos} \cite{ho2021integration}.

Neural Radiance Fields (NeRF) \cite{mildenhall2021nerf} have gained widespread attention in 3D vision and graphics due to their ability to generate novel views from sparse 2D inputs. By learning implicit 3D representations from a collection of images, NeRF maps 5D input coordinates — comprising a 3D position and a 2D viewing direction — to scene properties such as color and density. 
\begin{figure}[h]
    \centering
    \begin{subfigure}[b]{0.32\linewidth}
         \centering
            \includegraphics[trim={5cm 2cm 5cm 3cm},clip,width=0.45\linewidth]{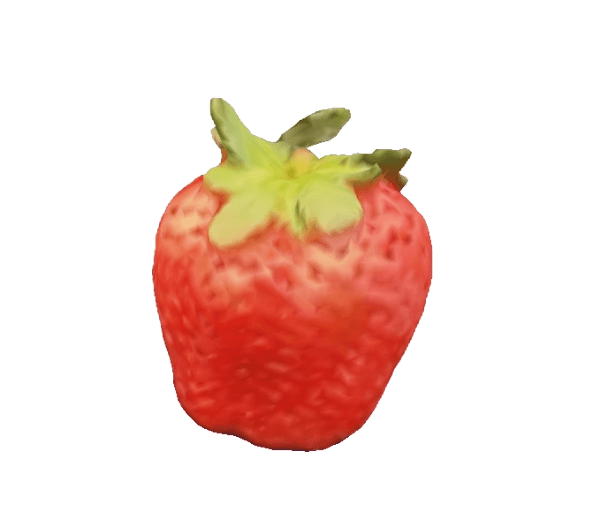}
            \includegraphics[trim={5cm 2cm 5cm 3cm},clip,width=0.45\linewidth]{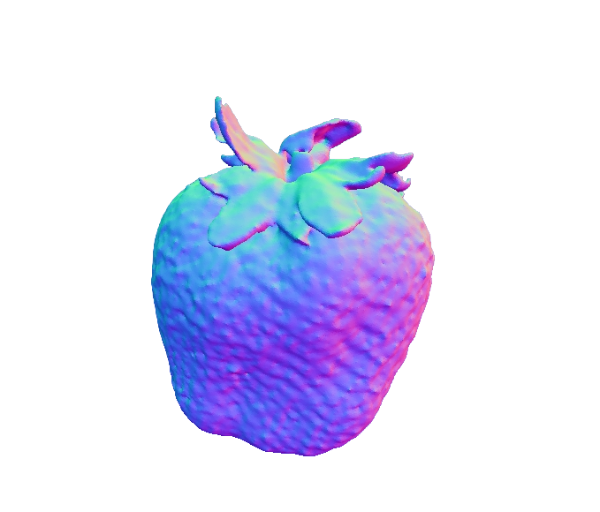}
         \caption{strawberry}
         \label{fig:rsl_1}
     \end{subfigure}
     \begin{subfigure}[b]{0.32\linewidth}
         \centering
            \includegraphics[trim={5cm 3cm 5cm 3cm},clip,width=0.45\linewidth]{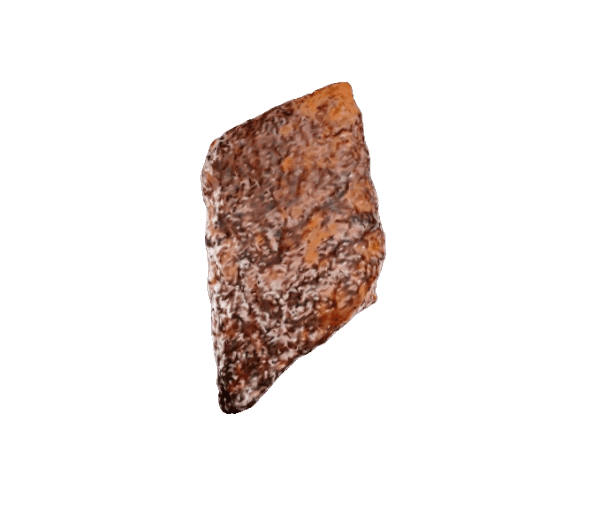}
            \includegraphics[trim={5cm 1cm 5cm 3cm},clip,width=0.45\linewidth]{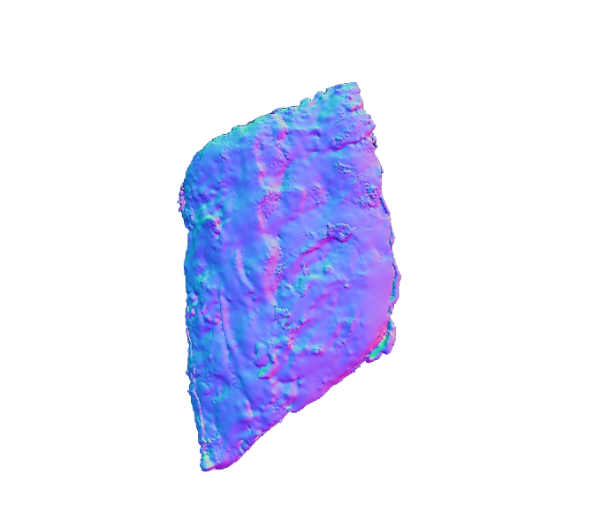}
         \caption{pork rib}
         \label{fig:rsl_3}
     \end{subfigure}
      \begin{subfigure}[b]{0.32\linewidth}
         \centering
            \includegraphics[trim={5cm 3cm 5cm 3cm},clip,width=0.45\linewidth]{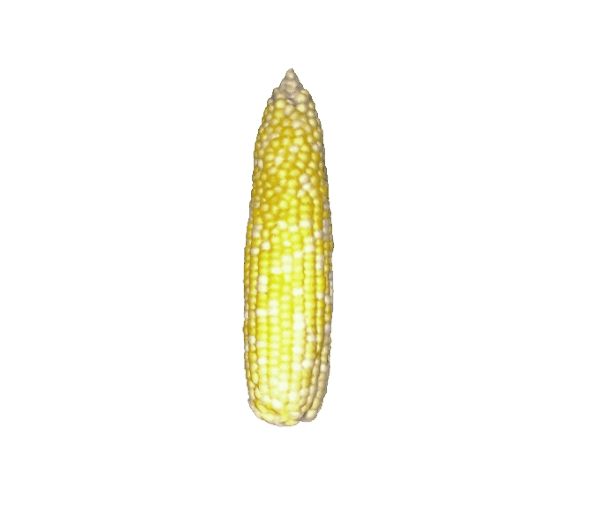}
            \includegraphics[trim={5cm 3cm 5cm 3cm},clip,width=0.45\linewidth]{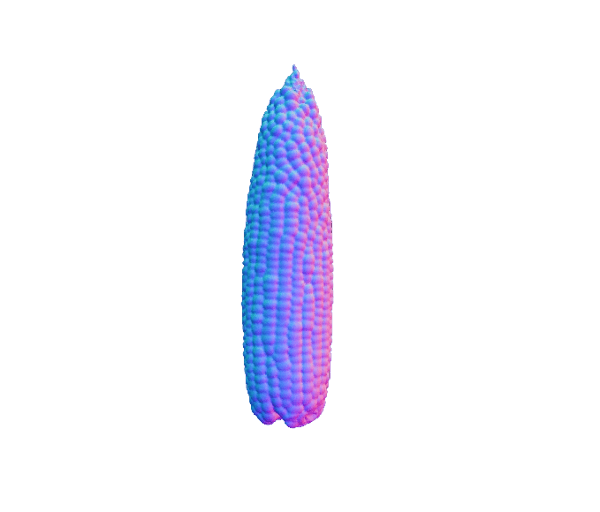}
         \caption{corn}
         \label{fig:rsl_4}
     \end{subfigure}
     
     \begin{subfigure}[b]{0.32\linewidth}
         \centering
            \includegraphics[trim={5cm 4cm 5cm 5cm},clip,width=0.45\linewidth]{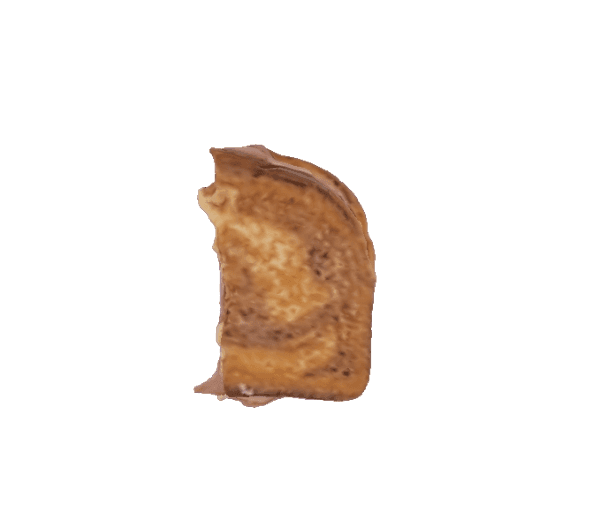}
            \includegraphics[trim={5cm 4cm 5cm 5cm},clip,width=0.45\linewidth]{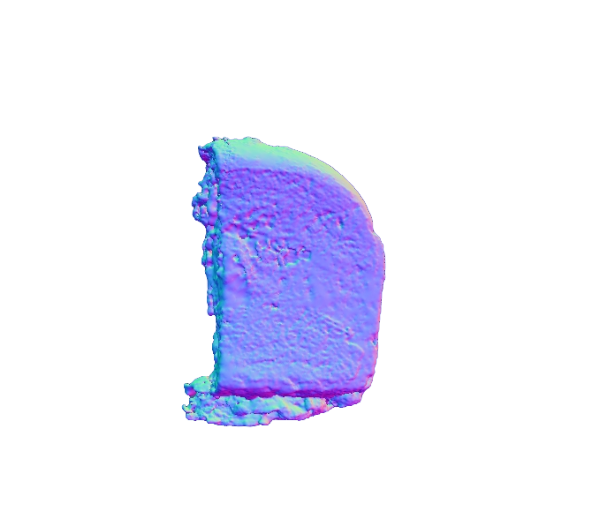}
         \caption{sandwich}
         \label{fig:rsl_6}
     \end{subfigure}
     \begin{subfigure}[b]{0.32\linewidth}
         \centering
            \includegraphics[trim={6cm 6cm 6cm 6cm},clip,width=0.45\linewidth]{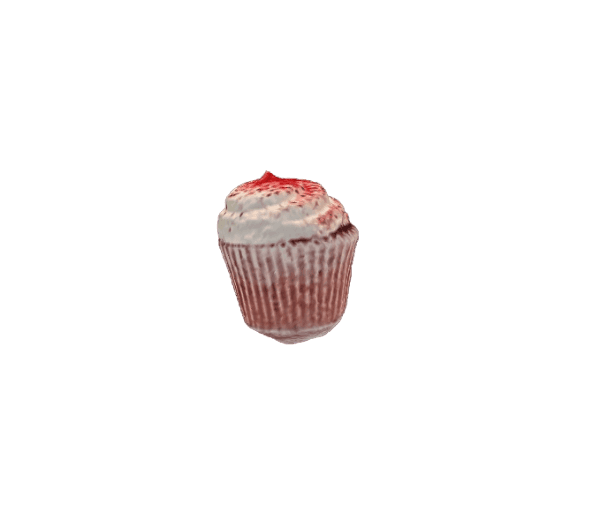}
            \includegraphics[trim={6cm 6cm 6cm 6cm},clip,width=0.45\linewidth]{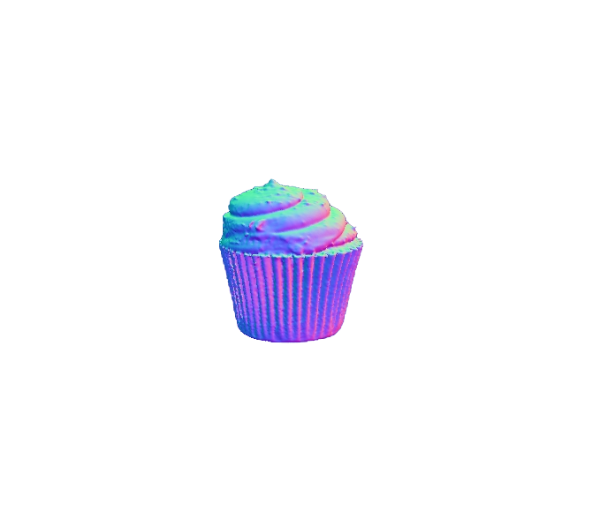}
         \caption{cake}
         \label{fig:rsl_8}
     \end{subfigure}
     \begin{subfigure}[b]{0.32\linewidth}
         \centering
            \includegraphics[trim={5cm 2cm 5cm 3cm},clip,width=0.45\linewidth]{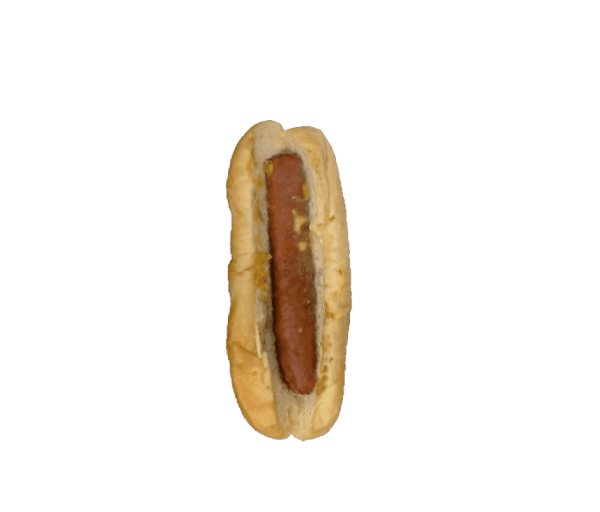}
            \includegraphics[trim={5cm 2cm 5cm 3cm},clip,width=0.45\linewidth]{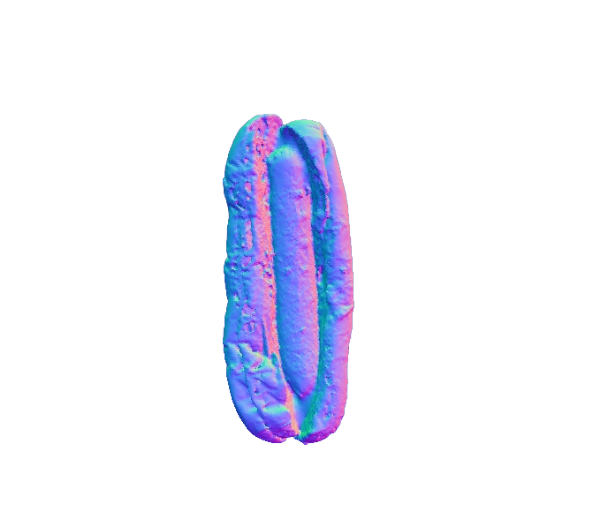}
         \caption{Hotdog}
         \label{fig:rsl_14}
     \end{subfigure}
    \caption{
    Comparison between our reconstruction (left) and the ground-truth (right) using the MTF dataset \cite{he2024metafood}. 
    }
    \label{fig:qualitative_results}
\end{figure}
This allows for accurate volume rendering and 3D reconstruction. Compared to direct estimation methods using 2D images, NeRF offers superior spatial fidelity, making it a promising approach for food volume estimation.

However, applying NeRF to food volume estimation presents several challenges. One major limitation is the lack of user control over the segmentation process. Existing NeRF-based approaches \cite{almughrabi2024voleta, he2024metafood}, focus on reconstructing food objects, but do not allow users to specify which food item should be segmented and analyzed. This limits the applicability of these methods in real-world scenarios where multiple food items may appear in a single scene, requiring selective processing. Additionally, NeRF methods require accurate object masks for object-centric reconstructions, but current solutions lack a mechanism for user-guided segmentation.

To address these challenges, we leverage Decoupled Video Segmentation \cite{cheng2023tracking}, a tracking-by-detection or a text-guided
segmentation, into our framework. Tracking-by-detection methods \cite{kim2015multiple, tang2017multiple, bergmann2019tracking, cheng2023tracking, almughrabi2025foodmem} rely on per-image detections and associate objects using short-term temporal tracking models. However, these approaches are highly dependent on the quality of individual frame detections and are prone to image-level errors. Although there are some long-term temporal propagation methods \cite{goel2021msn, garg2021mask}, they are typically designed for specific tasks and do not refine image-level segmentation. In contrast, DEVA \cite{cheng2023tracking}  introduces a bi-directional propagation mechanism that not only associates objects over time, but also denoises image-level segmentations, leading to more robust and consistent results across frames. This allows for improved and fast segmentation accuracy.

NeRF-based methods also depend on accurate camera localization to achieve high-quality 3D reconstructions \cite{mildenhall2021nerf}. Structure-from-motion (SfM) techniques such as COLMAP \cite{schonberger2016structure} provide robust camera pose estimation and sparse 3D reconstructions. Further enhancements like PixSfM \cite{lindenberger2021pixel} integrate deep learning techniques to improve pose estimation and feature matching, especially in challenging scenarios with limited views \cite{detone2018superpoint, sarlin2020superglue, lindenberger2021pixel}. By incorporating these methods, food object reconstructions become more reliable, even when the number of input images is minimal.

In this paper, we introduce VolTex, an enhanced human-in-the-loop framework that improves user control over food volume estimation by text-guided, decoupled video segmentation into our volume estimation framework. The remainder of this paper is structured as follows: Sec.~\ref{sec:related_work} presents related work, Sec.~\ref{sec:methodology} details our methodology, Sec.~\ref{sec:results} presents our qualitative and quantitative results, and Sec.~\ref{sec:conclusion} discusses limitations, conclusions, and future research directions.
\section{Related work}
\label{sec:related_work}
Food volume estimation has been extensively explored through 3D reconstruction techniques, leveraging advancements in structure-from-motion (SfM) \cite{schonberger2016structure}, NeRF \cite{mildenhall2021nerf}, and mesh extraction methods. These approaches integrate pose estimation, segmentation, and refinement techniques to improve accuracy and applicability.

Recent studies have introduced SfM-based pipelines to facilitate the reconstruction of food objects from multiple image viewpoints. The ININ. framework \cite{he2024metafood} provides a comprehensive 3D reconstruction solution that integrates COLMAP \cite{schonberger2016structure} for pose estimation, DiffusioNeRF \cite{wynn2023diffusionerf} for neural rendering regularization, and NeRF2Mesh \cite{tang2023delicate} for mesh extraction. To enhance the final meshes, this pipeline incorporates techniques for hole-filling and noise reduction, specifically Laplacian smoothing. In a similar vein, FoodR. \cite{he2024metafood} accomplishes high-quality food mesh reconstruction utilizing 3D Gaussian splatting \cite{kerbl20233d}, employing SfM for camera pose estimation alongside a dedicated post-processing pipeline aimed at scaling calibration and refinement. The methodologies employed facilitate detailed 3D reconstruction of food objects, thereby ensuring structural accuracy.

Neural implicit representations, particularly those based on NeRF, have significantly advanced the estimation of food volume by enabling dense three-dimensional reconstruction from limited input views. The VolETA framework 
\cite{almughrabi2024voleta, he2024metafood} integrates neural surface reconstruction alongside RGBD images and segmentation masks to facilitate accurate food volume estimation. This framework utilizes keyframe selection, perceptual hashing, and blur detection techniques to uphold input quality. Methods of neural surface reconstruction enhance the geometry of objects, while adjustments to the scaling factors ensure precise volume calculations. This approach establishes a comprehensive pipeline that merges SfM-based camera localization with implicit three-dimensional modeling, thereby improving its adaptability to real-world food estimation challenges.

While existing reconstruction pipelines prioritize the enhancement of mesh quality and pose estimation, they frequently operate under the assumption of automated segmentation without any user intervention. In practical scenarios involving the estimation of food volumes, multiple food items may be present within the same scene, thereby necessitating selective segmentation to achieve precise volume calculations. We \change{leverage} Decoupled Video Segmentation (DEVA) \cite{cheng2023tracking} to solve this challenge. 
Conventional tracking-by-detection methods \cite{kim2015multiple, tang2017multiple, bergmann2019tracking, cheng2023tracking} associate objects across frames through the utilization of short-term tracking models. DEVA improves segmentation accuracy by integrating a bi-directional propagation mechanism, which refines segmentation masks over time and mitigates inconsistencies across frames. This approach facilitates user-specified food segmentation, thereby ensuring that volume estimation is focused exclusively on the intended objects within a scene.

This paper presents VolTex, an innovative framework that amalgamates user-controlled segmentation with NeRF-based three-dimensional reconstruction to estimate food volume. The primary contributions of this study include:
\begin{itemize}

    \item 
    \change{We propose a text-guided food object volume estimation, showing the first exploration for user prompts in food volume estimation. This allows users to specify food items by name using text prompts for automated segmentation of a particular food object. Notably, it also automates reference object segmentation, enhancing accuracy and efficiency over VolETA.}
    \item We leverage implicit 3D reconstruction \change{of selected food objects} to estimate food volume while maintaining spatial consistency and high fidelity to reconstruction.
    \item  We conducted extensive experiments on the MTF dataset, evaluating segmentation quality, reconstruction accuracy, and overall volume estimation performance in real-world settings.
\end{itemize}
\section{Our Proposal: VolTex}
\label{sec:methodology}
Our framework employs multi-view reconstruction to generate detailed food meshes and accurately measure food volumes. Our framework is predicated on a multi-view reconstruction methodology designed to produce detailed food meshes and accurately estimate food volumes. In this section, we will initially provide an overview of our approach, including a discussion of the parameter extraction phases. Subsequently, we will elucidate the phases of Mesh Reconstruction and Volume Estimation within our model.

\begin{figure*}[ht]
    \centering
    \includegraphics[trim={4.5cm 0 1cm 5cm},clip,width=1\linewidth]{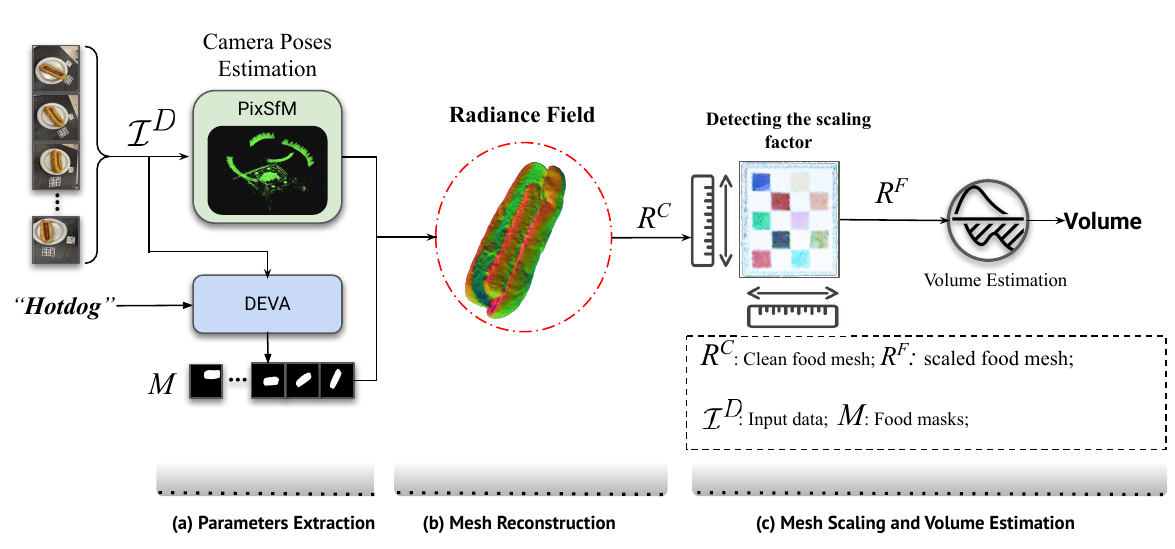}
    \caption{
Our approach for estimating food volume in  (a) \textbf{Parameters Extraction} uses PixSfM to estimate camera poses. Simultaneously, our approach segments the reference and food objects using DEVA with the text ``chessboard" and food label (provided by the dataset as shown in Table.~\ref{table:quntitative_results}) and images. After that, our approach applies a binary image segmentation method to images, reference object masks, and food object masks, resulting in RGBA images. In contrast, our approach transforms the RGBA images and poses to generate meaningful metadata and create modeled data. Next, (b) in \textbf{Mesh reconstruction}, we input the modeled data into NeuS2 to reconstruct colorful meshes for reference and food objects. To ensure accuracy, \change{our framework uses} ``Remove Isolated Pieces" with diameter thresholding to clean up the mesh and remove small isolated pieces that do not belong to the reference or food mesh resulting (\(\{R^C\}\)). Finally, we leverage the VolETA \cite{almughrabi2024voleta} scaling factors approach using the reference mesh via MeshLab (\(S\)). We fine-tune the scaling factor using depth information and the food masks and then apply the fine-tuned scaling factor (\(S_f\)) to the cleaned food mesh to generate a scaled food mesh (\(R^F\)) in meter units.
\change{We also use the food labels ``Strawberry", ``Cinnamon bun", ``Pork rib", ``Corn", ``French toast", ``Sandwich", ``Burger", ``Cake", ``Blueberry muffin", "Banana", ``Salmon", and ``Burrito" to segment different food objects.}  
}
\label{fig:methodology}
\end{figure*}

\subsection{Overview}
Our approach, \textbf{VolTex}, reconstructs 3D food meshes from multiple images taken from different viewpoints and estimates their volume using reference-based scaling. The pipeline consists of three key phases:  Parameters Extraction in Sec.~\ref{sec:parameter_extraction}, Mesh Reconstruction in Sec.~\ref{sec:mesh_reconstruction}, and Volume Estimation in Sec.~\ref{sec:volume_estimation}. Given a sequence of input images and corresponding food labels, we first estimate camera poses and segmentation masks, reconstruct food and reference meshes, and finally compute the real-world volume by applying a scaling factor derived from a reference object.

\subsection{Methodology}
\label{sec:methodology}
Our pipeline consists of three key phases: (a) Parameters Extraction and text-guided video segmentation, (b) Mesh Reconstruction, and (c) Volume Estimation, as shown in Fig.~\ref{fig:methodology}.

\subsubsection{Parameters Extraction and Text-guided Video Segmentation}
\label{sec:parameter_extraction}
For an accurate 3D reconstruction, we estimate camera parameters and segment the food and reference objects.
Given a set of input images \( \mathcal{I} = \{\mathcal{I}_i\}_{i=1}^{n} \), we estimate the camera intrinsics and extrinsics \( C = \{C_j\}_{j=1}^{n} \) using \textbf{PixSfM}~\cite{lindenberger2021pixel}, an advanced Structure-from-Motion (SfM) pipeline. PixSfM enhances traditional COLMAP-based SfM by integrating SuperPoint feature extraction and SuperGlue graph-based matching, improving pose estimation in complex and low-texture scenes.

\change{DEVA \cite{cheng2023tracking} constitutes a decoupled segmentation framework that integrates an image segmentation model with a universal temporal propagation model to achieve superior video object segmentation. The image segmentation model, which is trained on task-specific data, produces segmentation hypotheses at the frame level. Concurrently, the temporal propagation model, developed using class-agnostic mask propagation datasets, links and disseminates these hypotheses throughout the entire video. This decoupled architecture enables DEVA to exhibit robust generalization capabilities, particularly in scenarios, where labeled training data is limited.} For \change{the different food objects segmentation}, we leverage \textbf{DEVA} \cite{cheng2023tracking}, enabling user-guided object segmentation. The user provides a reference label (e.g. ``chessboard") and a \textit{food label} (from the dataset, as shown in Table.~\ref{table:quntitative_results}), producing a set of segmentation masks \( \{M^F_i, M^R_i\}_{i=1}^{n} \), where \( M^F \) represents the food mask, and \( M^R \) represents the reference object mask. These masks are applied to the images, isolating the food and reference objects from the background. This step ensures that only relevant regions corresponding to the preselected objects in the images contribute to the 3D reconstruction process, as shown in Fig.~\ref{fig:methodology} (a).

\subsubsection{Mesh Reconstruction}
\label{sec:mesh_reconstruction}
Once the segmented images and camera poses are obtained, we reconstruct the 3D food mesh using Neural Surface Reconstruction. We employ \textbf{NeuS2}~\cite{wang2023neus2} to model the object's surface as a signed distance function (SDF), where the reconstructed implicit surface is defined as:
\begin{equation}
    \mathcal{S} = \{ \mathbf{x} \in \mathbb{R}^3 \mid f(\mathbf{x}) = 0 \},
\end{equation}
where \( f(\mathbf{x}) \) maps each point in \( \mathbb{R}^3 \) to its signed distance from the object boundary. Given camera rays \( \mathbf{r}(t) = \mathbf{o} + t\mathbf{d} \), NeuS2 optimizes a volume rendering function:
\begin{equation}
    \mathbf{C}(\mathbf{r}) = \int_{t_n}^{t_f} T(t) \sigma(\mathbf{r}(t)) \mathbf{c}(\mathbf{r}(t)) \, dt,
\end{equation}
where \( T(t) \) represents transmittance, and \( \sigma(\mathbf{r}(t)) \) denotes the density at each point along the ray. The final 3D reconstruction consists of the \textbf{food mesh \( R^F \) and reference mesh \( R^R \)}:
\begin{equation}
    R = \{R^F, R^R\}.
\end{equation}
To refine the food mesh, we apply the ``Remove Isolated Pieces" technique from MeshLab~\cite{cignoni2008meshlab}. Given that our scenes contain only one food item, we measure the \textbf{diameter of each isolated mesh piece} and compare it to the \textbf{largest piece in \( R^F \)}. Any isolated mesh piece with a diameter less than 5\% of the largest piece is removed. The resulting cleaned food mesh is defined as:
\begin{equation}
    R^C = \{r \in R^F \mid \operatorname{diam}(r) > \delta \times \operatorname{diam}(R^F_{\max})\},
\end{equation}
where \( R^F_{\max} \) represents the largest connected food mesh component. $\delta$ is the diameter threshold constant. Notably, no refinement is applied to the reference mesh; it remains unchanged for scaling factor computation. This step ensures that small artifacts and noise are removed only from the food object, preserving an accurate reconstruction for volume estimation, as shown Fig.~\ref{fig:methodology} (b).

\subsubsection{Volume Estimation}
\label{sec:volume_estimation}
Once the cleaned food mesh and unchanged reference mesh are obtained, we compute the real-world food volume by applying a scaling factor to the unitless NeRF-based reconstruction. Since NeRF outputs unscaled meshes, we determine the \textbf{scaling factor \( s \)} using a reference object (e.g., a checkerboard) with a predefined dimension \( \ell_{\text{real}} \). The scaling factor is computed as:
\begin{equation}
    s = \frac{\ell_{\text{real}}}{\text{med}(d^j)},
\end{equation}
where \( d^j \) represents the \textbf{median Euclidean distance} between projected 3D corner points of the reference object. Applying the scaling factor to the food mesh results in:
\begin{equation}
    R^F_{\text{scaled}} = s \cdot R^F.
\end{equation}
Finally, we compute the \textbf{food object volume \( V_F \)} using \textbf{tetrahedral decomposition}:
\begin{equation}
    V_F = \sum_{i} \frac{1}{6} \left| \vec{a}_i \cdot (\vec{b}_i \times \vec{c}_i) \right|,
\end{equation}
where \( \vec{a}_i, \vec{b}_i, \vec{c}_i \) define the triangular faces of the 3D mesh, and the determinant computes the signed volume of each tetrahedral element, as shown Fig.~\ref{fig:methodology} (c).

\section{Validation}
\label{sec:results}
In this section, we begin with the implementation settings discussion presented in Sec.~\ref{sec:resource_limitation}. We describe the datasets in Sec.~\ref{sec:datasets}, then outline the evaluation protocol in Sec.~\ref{sec:evaluation_protocol}. Finally, Sec.~\ref{sec:voleta_results} presents the qualitative and quantitative results.
Our approach is evaluated on a 3D food dataset from the MetaFood CVPR workshop \cite{he2024metafood}.

\subsection{Implementation settings}
\label{sec:resource_limitation}
We ran the experiments using two GPUs, GeForce GTX 1080 Ti/12G and RTX 3060/6G. We set the $\delta$ diameter as 5\% of the mesh size for removing isolated pieces. The number of iteration of NeuS2 is 5000, mesh resolution is $256 \times 256$, the unit cube ``aabb\_scale" is $1$, ``scale": 0.15, and ``offset": $[0.5, 0.5, 0.5]$ for each food scene. 

\subsection{Datasets}
\label{sec:datasets}
\textbf{MTF Dataset} includes 20 food scenes categorized into easy (8 scenes, ~200 images each), medium (7 scenes, ~30 images each), and hard (1 image per scene). Each scene contains food masks, a reference board (e.g., chessboard), QR codes, and metadata describing the food objects. Although depth images are available, we exclude them to ensure real-world applicability\footnote{Note that we do not use these images in order to make the method applicable to real and generic conditions}. The dataset is publicly available in\footnote{https://www.kaggle.com/competitions/cvpr-metafood-3d-food-reconstruction-challenge/data}. 
\change{Notably, we leverage MetaFood3D, the sole dataset that offers meshes, volume annotations, and multi-view images for estimating food volume. With its high-quality images, depth data, and reference objects, it serves as the most appropriate benchmark for assessing segmentation, 3D reconstruction, and scaling factor estimation.}

\subsection{Evaluation Protocol}
\label{sec:evaluation_protocol}
The evaluation consists of two phases: portion size accuracy (volume) and shape accuracy (3D structure).  

\textbf{Phase I} measures portion size accuracy using the Mean Absolute Percentage Error (MAPE):  
\begin{equation}
    \text{MAPE} = \frac{1}{n} \sum_{i=1}^{n} \left| \frac{V_{\text{true}, i} - V_{\text{pred}, i}}{V_{\text{true}, i}} \right| \times 100\% 
\end{equation}
where \( V_{\text{true}, i} \) and \( V_{\text{pred}, i} \) are the true and predicted volumes for the \( i \)-th model.  

\textbf{Phase II} validates volume accuracy by evaluating the complete 3D mesh using the Chamfer distance \cite{barrow1977parametric}, which quantifies shape similarity between the reconstructed models and the ground truth. This two-phase evaluation ensures a comprehensive assessment of both portion size estimation and structural accuracy.

\subsection{\change{VolTex} Results}
\label{sec:voleta_results}

\begin{table*}[h]
\footnotesize
\centering
\setlength{\tabcolsep}{0.2em}

\caption{Predicted volumes (in cubic millimeters) for various food items by Ours, VolETA \cite{almughrabi2024voleta,he2024metafood}, ININ. \cite{he2024metafood}, and FoodR. \cite{he2024metafood}, alongside ground truth volumes. The MAPE and Chamfer distances are also reported for each method. $\star$The relative mean error decreases when applying our method.}

\label{tab:volume_comparison}
    \label{table:quntitative_results}
    \begin{tabular}{c|l|cccc|c|cccc|cccc}
    \hline
    \multirow{2}{*}{ID} & \multirow{2}{*}{Label} & \multicolumn{4}{c|}{Absolute Error↓} & \multirow{2}{*}{Volume} & \multicolumn{4}{c|}{APE (\%)↓} & \multicolumn{4}{c}{Ch. Dis. (w/ transforms)↓}  \\ 

     & & \textbf{Ours} & VolETA & ININ. & FoodR. & & \textbf{Ours} & VolETA   & ININ.  & FoodR.  & \textbf{Ours} & VolETA   & ININ.  & FoodR.   \\ \hline
1 & Strawberry &  1.53 & 1.53 & 0.88 & 5.98 & 38.53 & 3.96 & 3.97 & 2.28 & 15.52 & 0.0015 & 0.0016 & 0.002 & 0.0011\\
2 & Cinnamon bun &  39.69 & 63.46 & 45.08 & 40.9 & 280.36 & 14.16 & 22.64 & 16.08 & 14.59 & 0.0051 & 0.0071 & 0.0036 & 0.0031\\
3 & Pork rib & 19.11 & 29.21 & 223.75 & 86.46 & 249.65 & 7.65 & 11.7 & 89.63 & 34.63 & 0.0144 & 0.0137 & 0.0049 & 0.0053\\
4 & Corn & 13.23 & 16.11 & 0.81 & 52.41 & 295.13 & 4.48 & 5.46 & 0.27 & 17.76 & 0.0025 & 0.0020 & 0.0038 & 0.0015\\
5 & French toast & 10.68 & 3.18 & 38.92 & 3.3 & 392.58 & 2.72 & 0.81 & 9.91 & 0.84 & 0.0114 & 0.0137 & 0.002 & 0.0040\\
6 & Sandwich & 3.39 & 13.14 & 19.57 & 20.49 & 218.31 & 1.55 & 6.02 & 8.96 & 9.39 & 0.0054 & 0.0067 & 0.0038 & 0.0025\\
7 & Burger & 8.25 & 4.16 & 7.28 & 43.75 & 368.77 & 2.24 & 1.13 & 1.97 & 11.86 & 0.0054 & 0.0047 & 0.0048 & 0.0025\\
8 & Cake & 5.61 & 13.49 & 0.81 & 8.08 & 173.13 & 3.24 & 7.79 & 0.47 & 4.67 & 0.0030 & 0.0030 & 0.0019 & 0.0010\\
9 & Blueberry muffin & 25.19 & 8.66 & 20.27 & 1.05 & 232.74 & 10.82 & 3.72 & 8.71 & 0.45 & 0.0059 & 0.0039 & 0.0029 & 0.0033\\
10 & Banana & 1.94 & 9.47 & 5.65 & 3.17 & 163.23 & 1.19 & 5.8 & 3.46 & 1.94 & 0.0025 & 0.0027 & 0.0034 & 0.0019\\
11 & Salmon & 2.28 & 4.78 & 8.72 & 0.82 & 85.18 & 2.67 & 5.61 & 10.24 & 0.96 & 0.0041 & 0.0034 & 0.0015 & 0.0015\\
13 & Burrito & 6.33 & 55.71 & 61.68 & 26.42 & 308.28 & 2.05 & 18.07 & 20.01 & 8.57 & 0.0042 & 0.0052 & 0.0026 & 0.0041\\
14 & Hotdog & 7.64 & 54.38 & 94.72 & 72.07 & 589.82 & 1.29 & 9.22 & 16.06 & 12.22 & 0.0047 & 0.0043 & 0.0044 & 0.0046\\ \hline
Mean & -& \textbf{11.14} & 21.33 & 40.63 & 28.069 & - & \textbf{4.46} & 7.84 & 14.47 & 10.26 & \textbf{0.0054} & 0.0055 & 0.0032 & 0.0028\\
Stdev. &- &  \textbf{11.07} & 22.10 & 61.84 & 28.87 & - & \textbf{4.01} & 6.36 & 23.47 & 9.48 & \textbf{0.0036} & 0.0040 & 0.0011 & 0.0014 \\ 
Rel.$\star$&- &  - & +91\% & +265\% & +152\% & - & - & +75\% & +218\% & +130\% & - & +19\% & -40\% & -48\% \\ \hline
    \end{tabular}
\end{table*}

\begin{figure*}[htb]
    \centering
    \footnotesize
    \setlength{\tabcolsep}{1pt}
    \begin{tabular}{c|c|c|c|c|c|c|c|c|c|c|c|c|c}
        \hline
          & 1 & 2 & 3 & 4 & 5 & 6 & 7 & 8 & 9 & 10 & 11 & 13 & 14         
        \\
        
        \hline
        \rotatebox{90}{Reference} &
        \includegraphics[clip,width=0.07\linewidth]{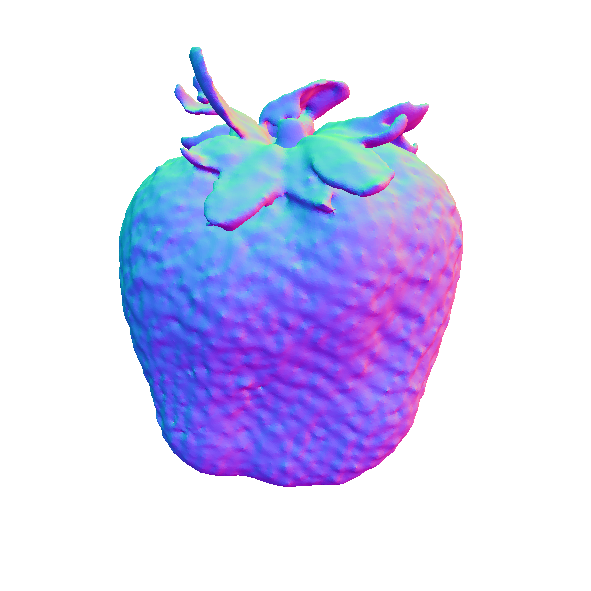} &
        \includegraphics[clip,width=0.07\linewidth]{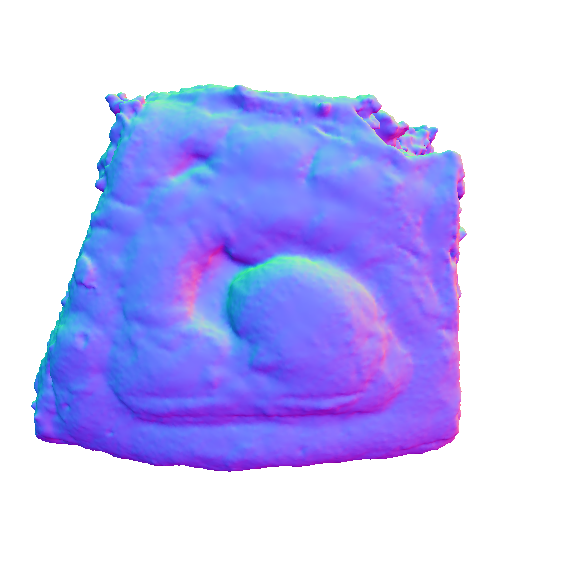} &
        \includegraphics[clip,width=0.07\linewidth]{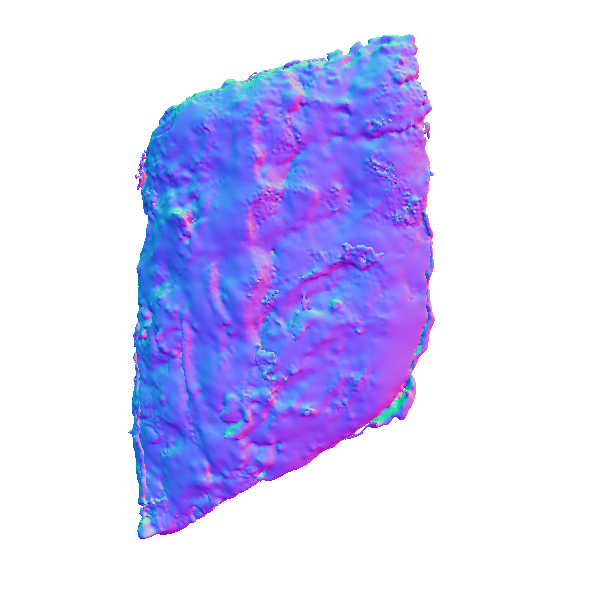} &
        \includegraphics[clip,width=0.07\linewidth]{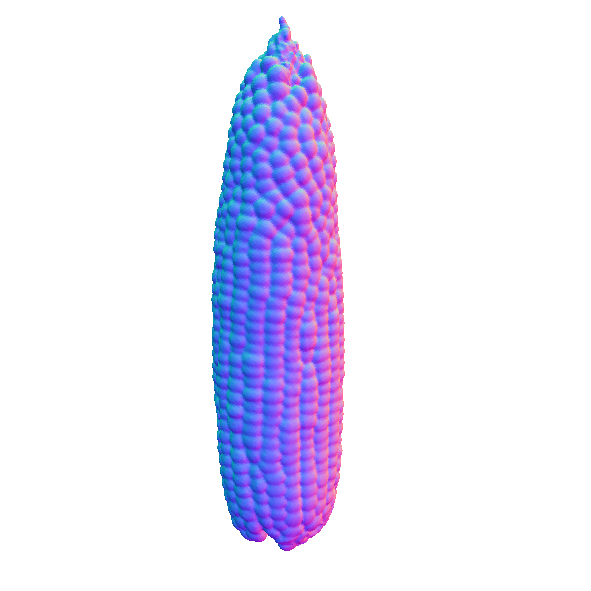} &
        \includegraphics[clip,width=0.07\linewidth]{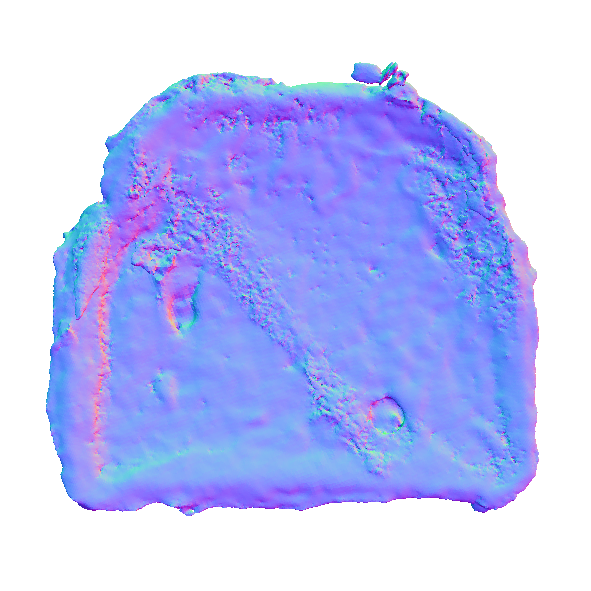} &
        \includegraphics[clip,width=0.07\linewidth]{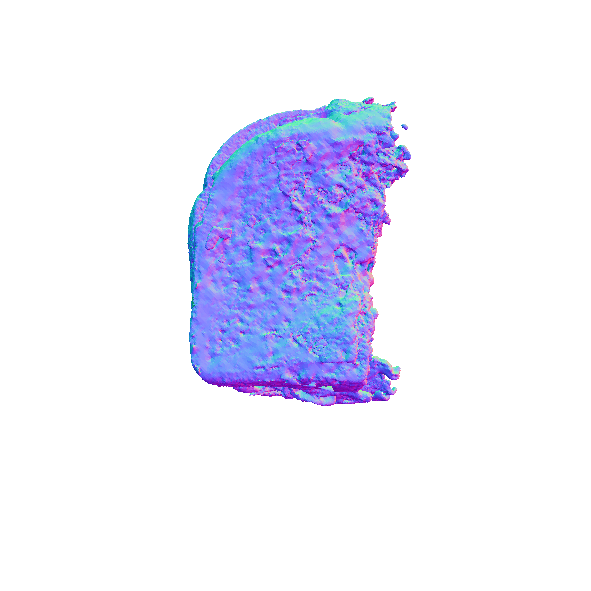} &
        \includegraphics[clip,width=0.07\linewidth]{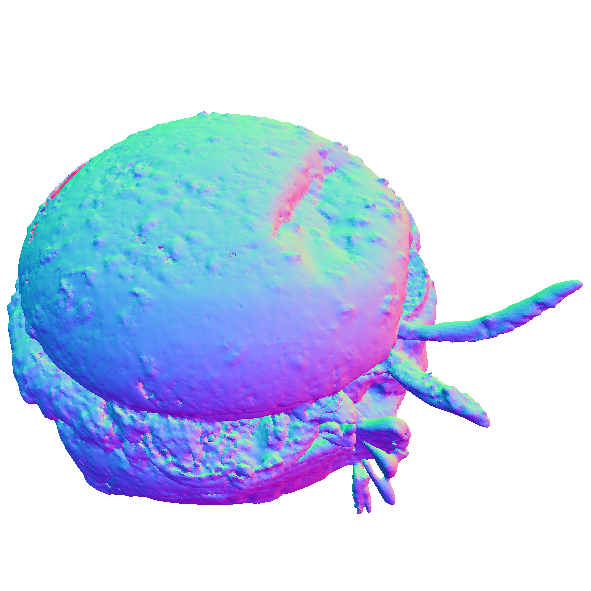} &
        \includegraphics[clip,width=0.07\linewidth]{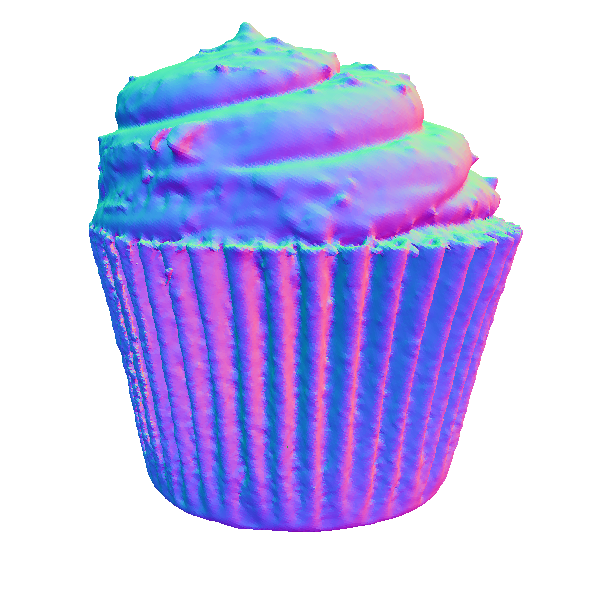} &
        \includegraphics[clip,width=0.07\linewidth]{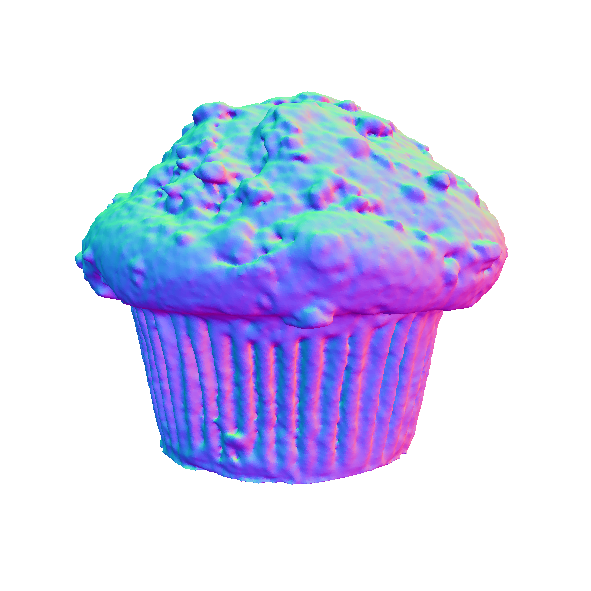} &
        \includegraphics[clip,width=0.07\linewidth]{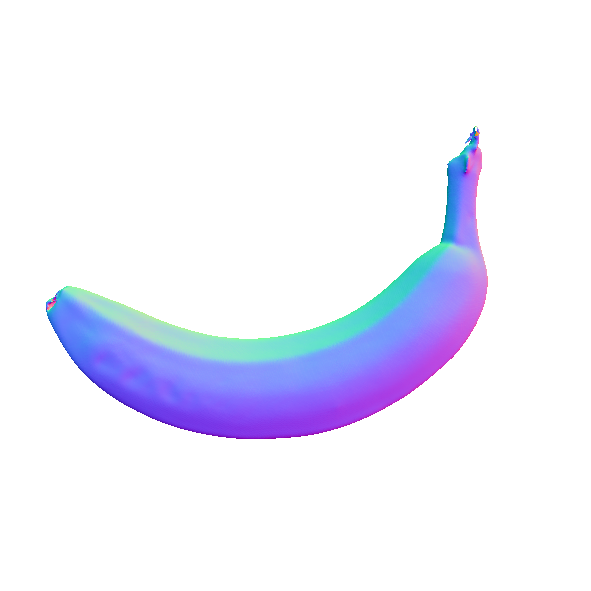} &
        \includegraphics[clip,width=0.07\linewidth]{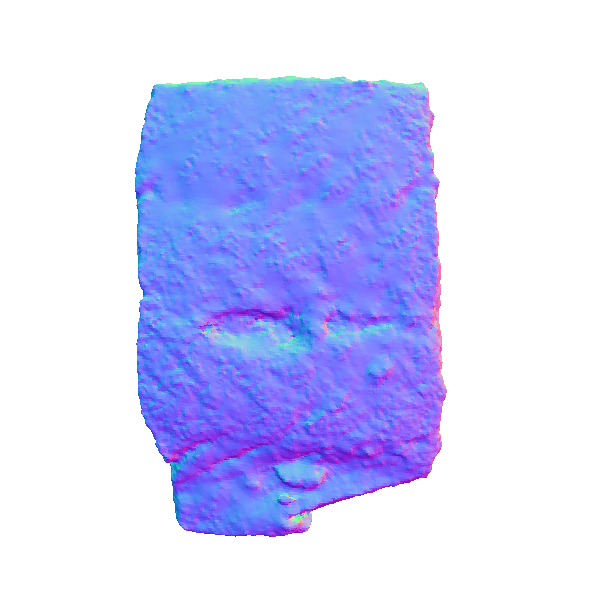} &
        \includegraphics[clip,width=0.07\linewidth]{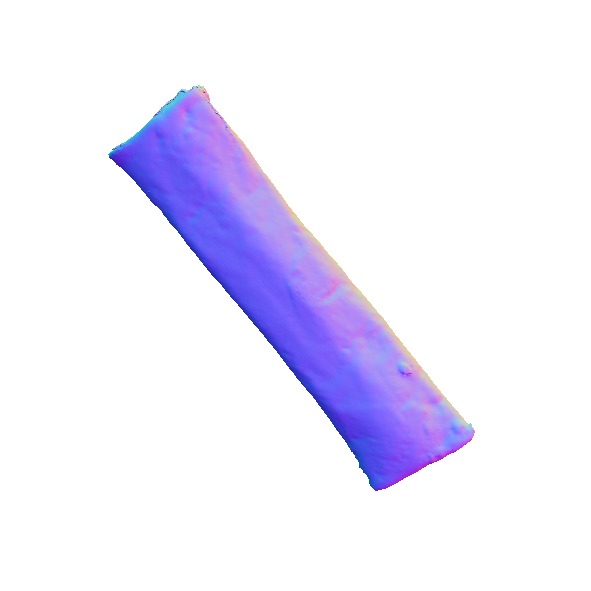} &
        \includegraphics[clip,width=0.07\linewidth]{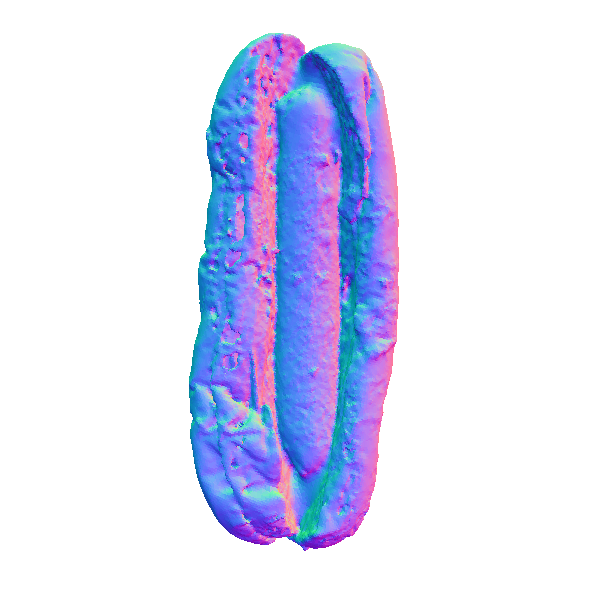}
        \\
          \rotatebox{90}{VolETA} &
        \includegraphics[clip,width=0.07\linewidth]{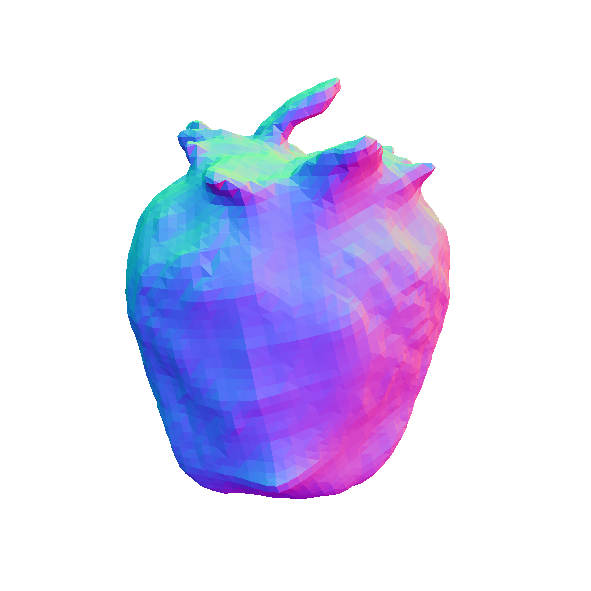} &
        \includegraphics[clip,width=0.07\linewidth]{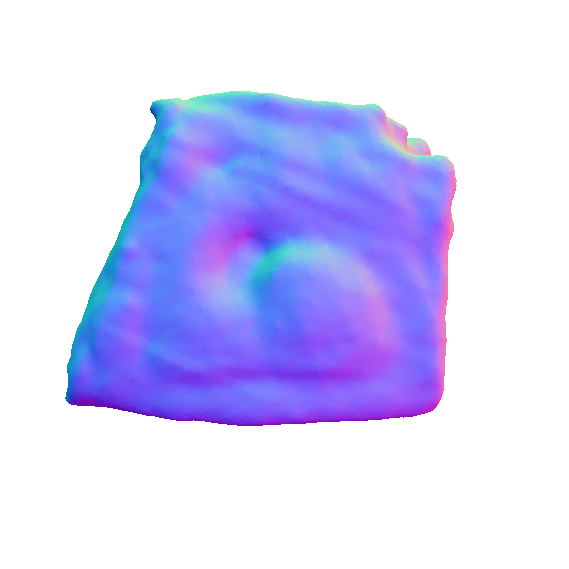} &
        \includegraphics[clip,width=0.07\linewidth]{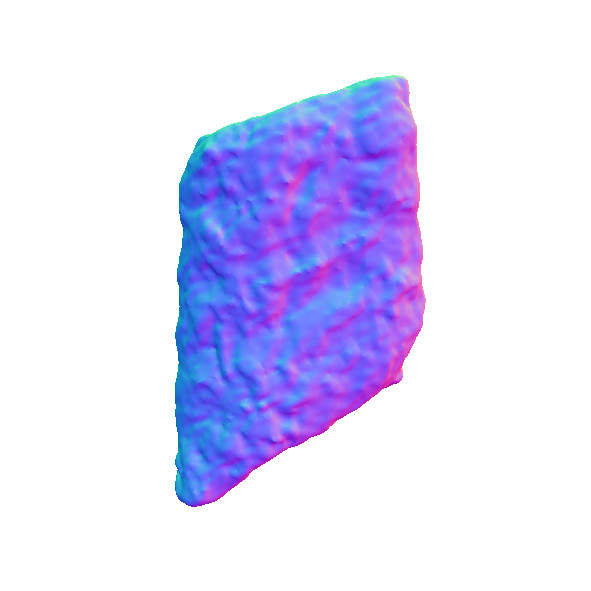} &
        \includegraphics[clip,width=0.07\linewidth]{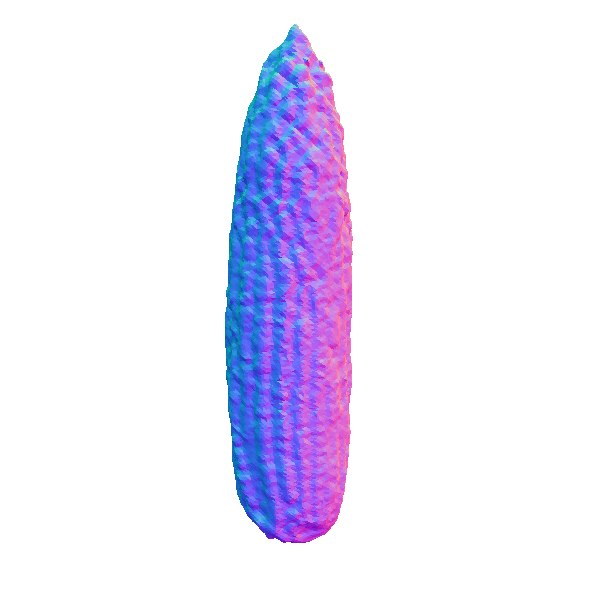} &
        \includegraphics[clip,width=0.07\linewidth]{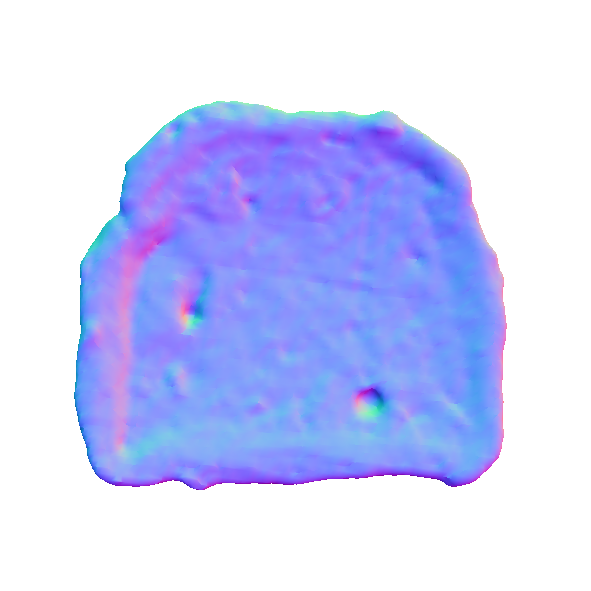} &
        \includegraphics[clip,width=0.07\linewidth]{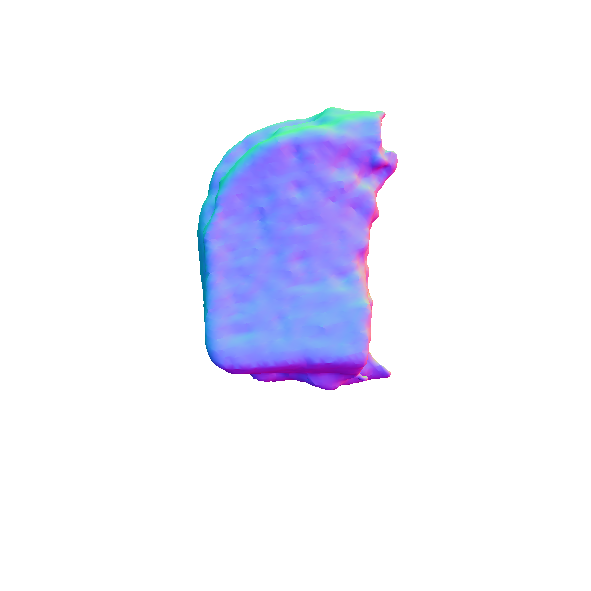} &
        \includegraphics[clip,width=0.07\linewidth]{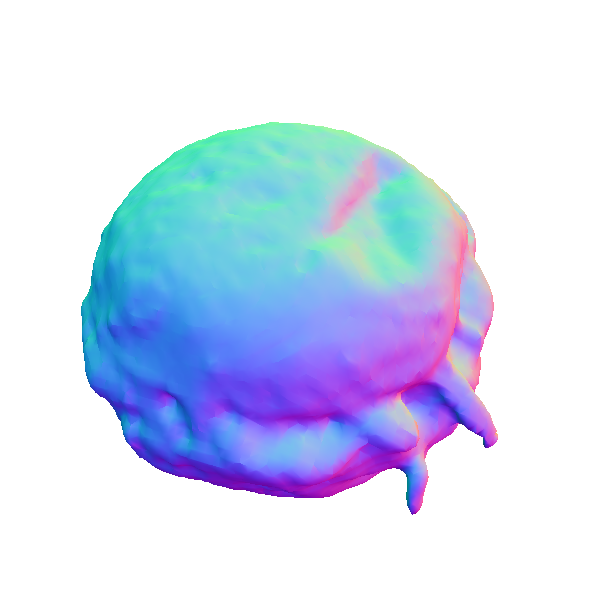} &
        \includegraphics[clip,width=0.07\linewidth]{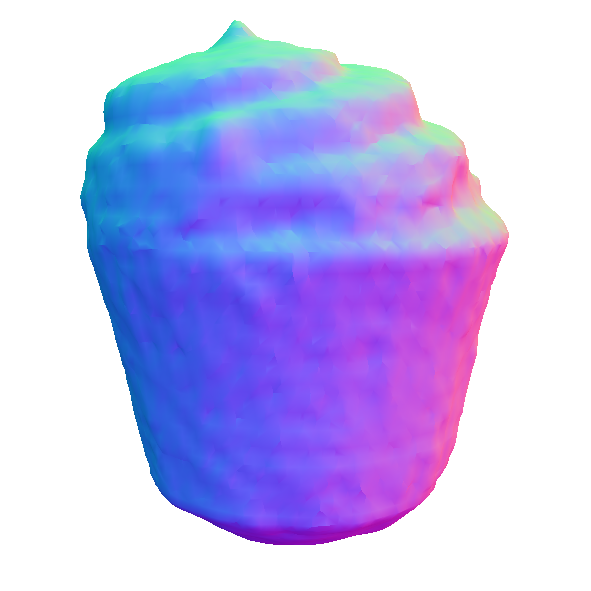} &
        \includegraphics[clip,width=0.07\linewidth]{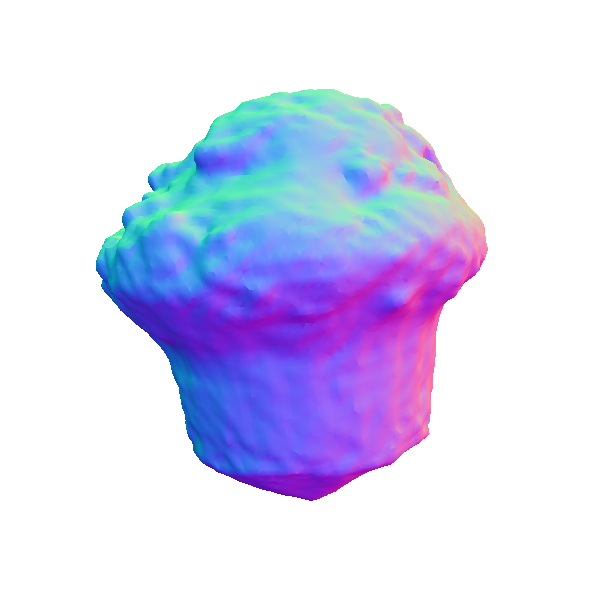} &
        \includegraphics[clip,width=0.07\linewidth]{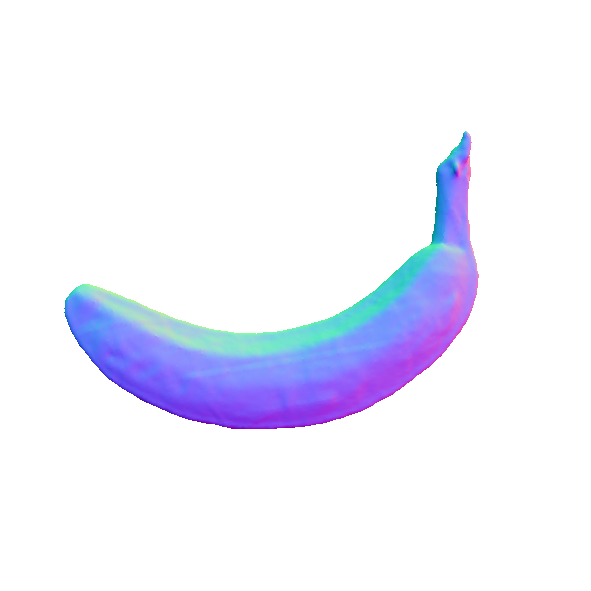} &
        \includegraphics[clip,width=0.07\linewidth]{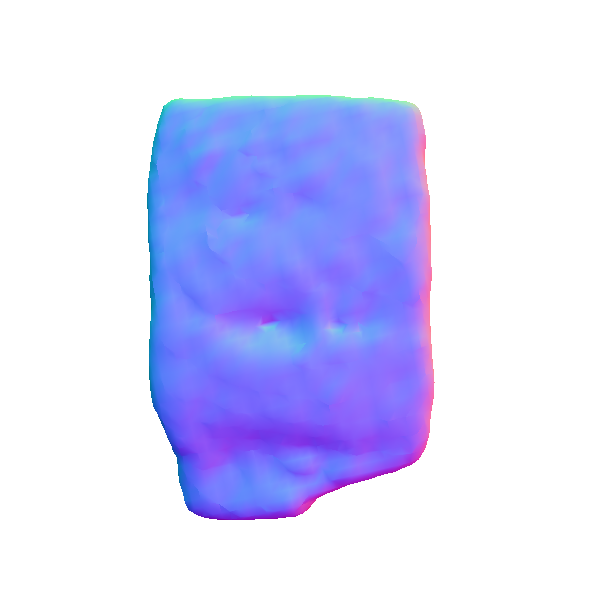} &
        \includegraphics[clip,width=0.07\linewidth]{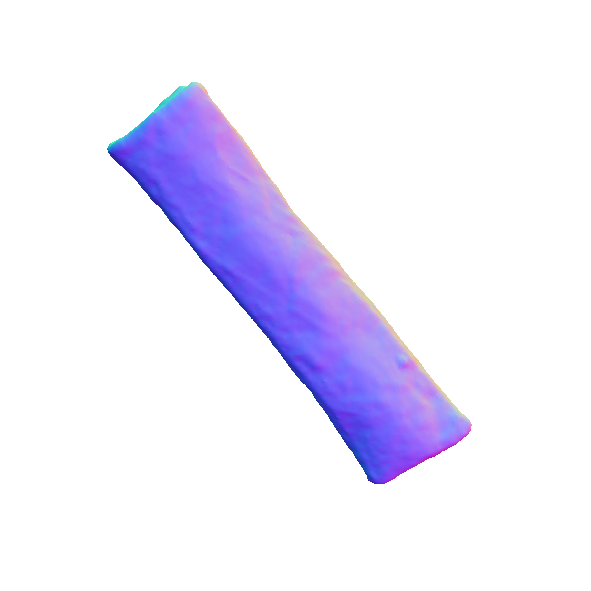} &
        \includegraphics[clip,width=0.07\linewidth]{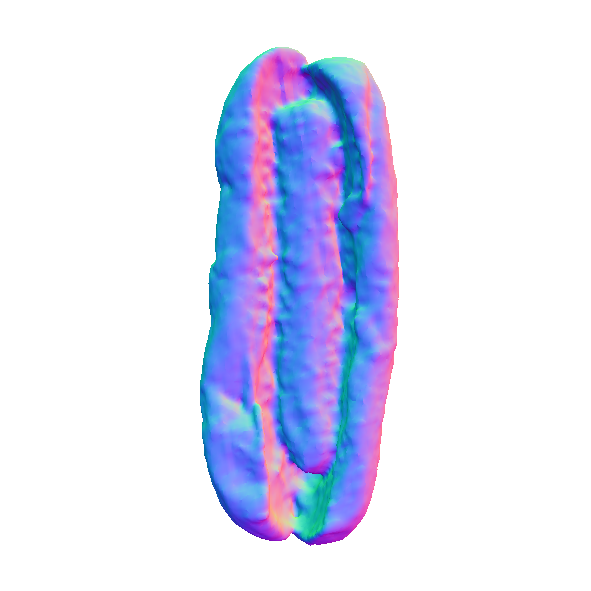}
        \\
        \rotatebox{90}{Ours} &
        \includegraphics[clip,width=0.07\linewidth]{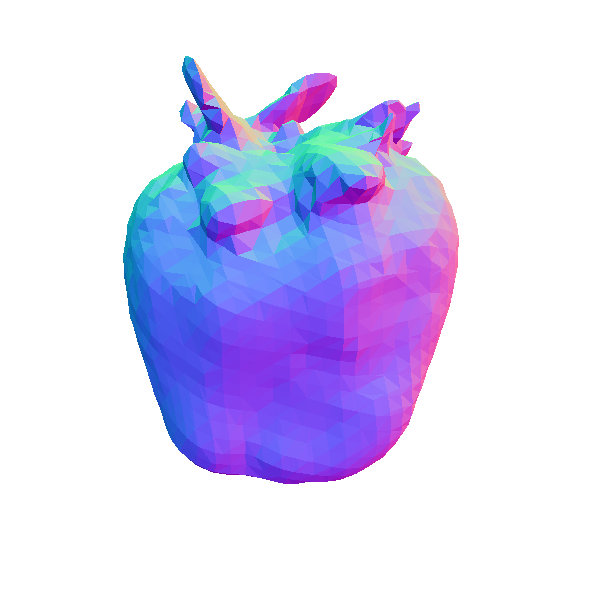} &
        \includegraphics[clip,width=0.07\linewidth]{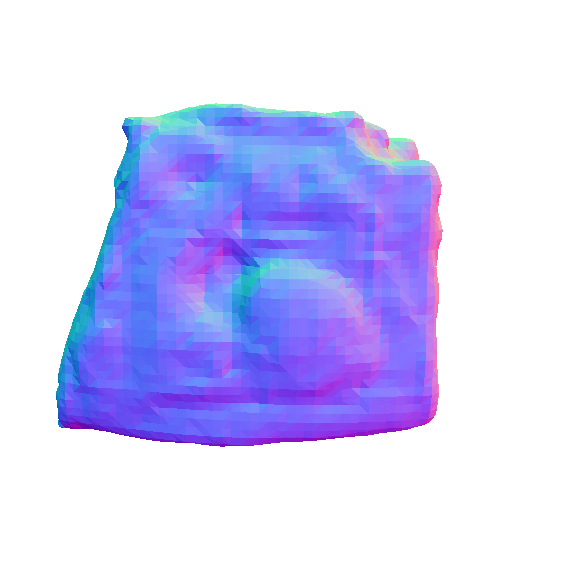} &
        \includegraphics[clip,width=0.07\linewidth]{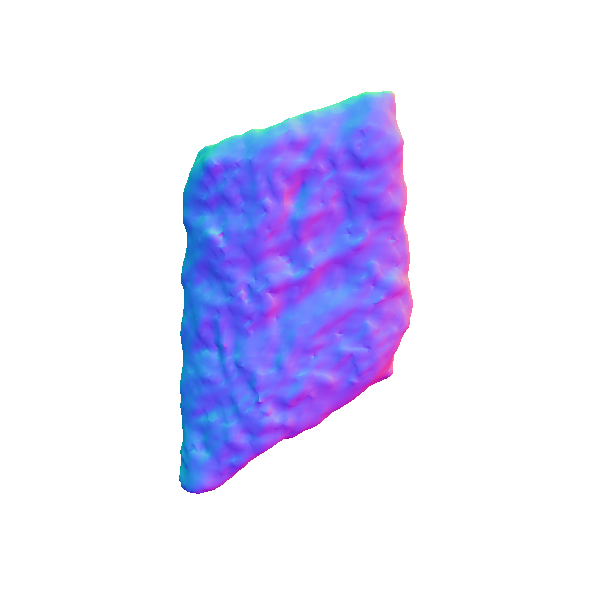} &
        \includegraphics[clip,width=0.07\linewidth]{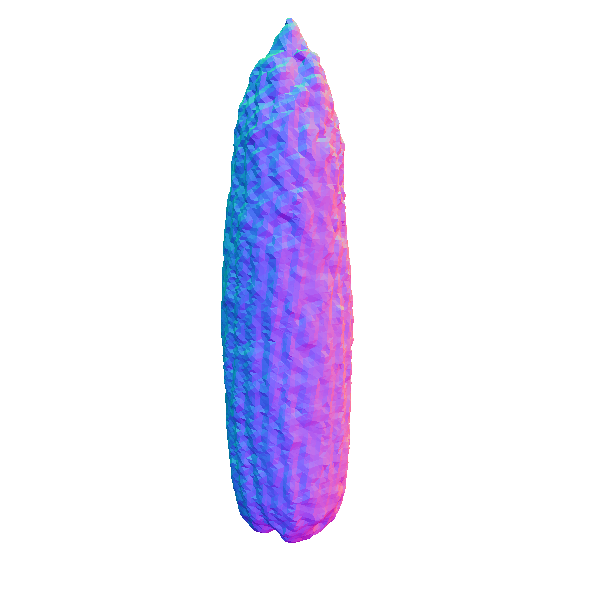} &
        \includegraphics[clip,width=0.07\linewidth]{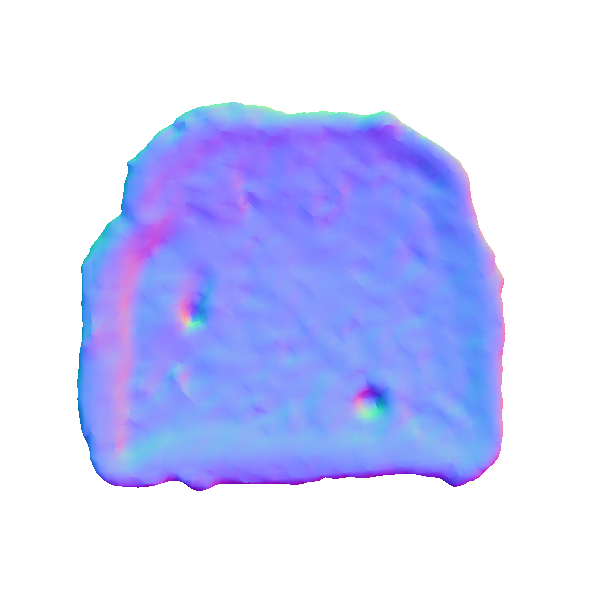} &
        \includegraphics[clip,width=0.07\linewidth]{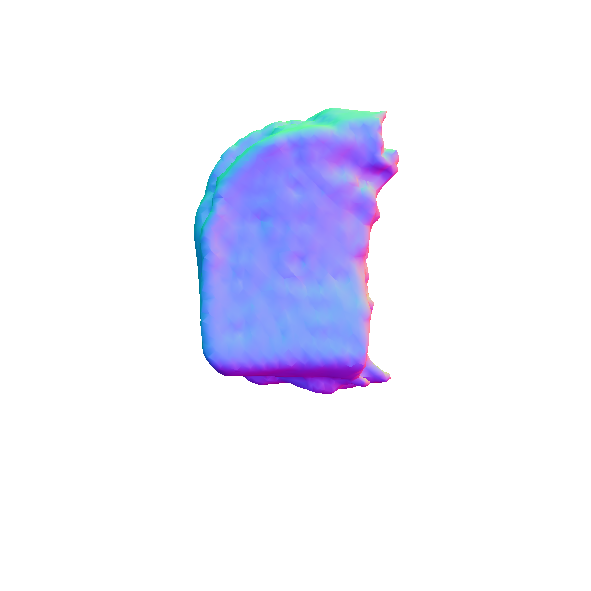} &
        \includegraphics[clip,width=0.07\linewidth]{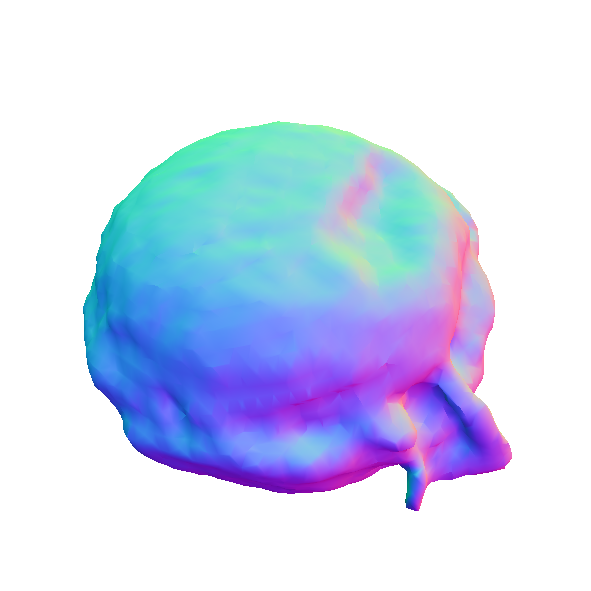} &
        \includegraphics[clip,width=0.07\linewidth]{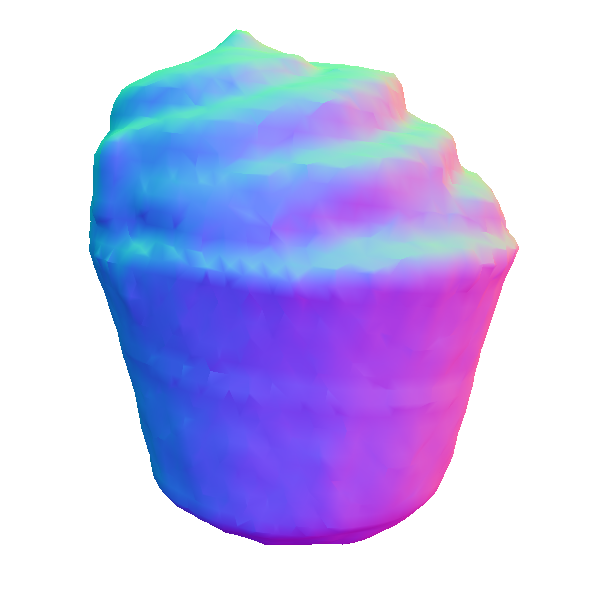} &
        \includegraphics[clip,width=0.07\linewidth]{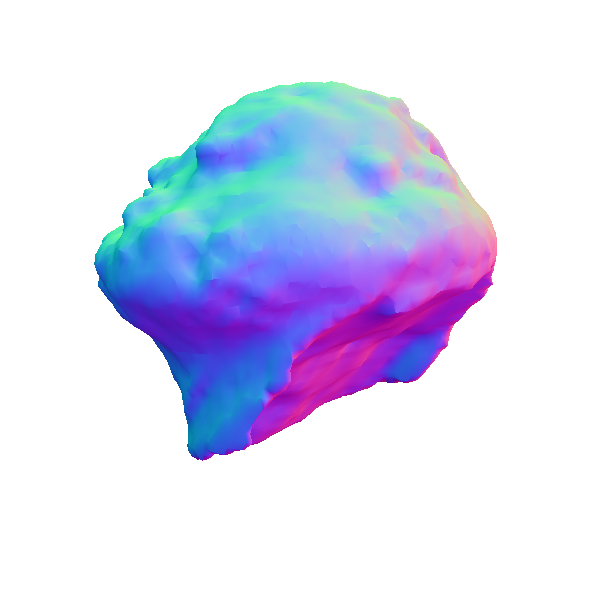} &
        \includegraphics[clip,width=0.07\linewidth]{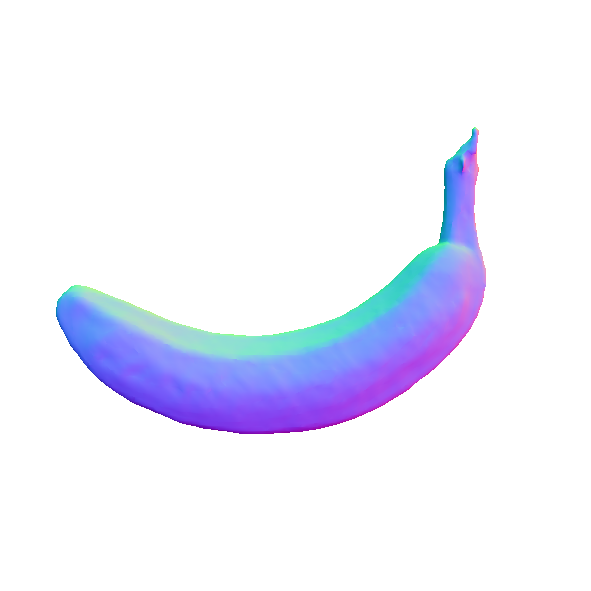} &
        \includegraphics[clip,width=0.07\linewidth]{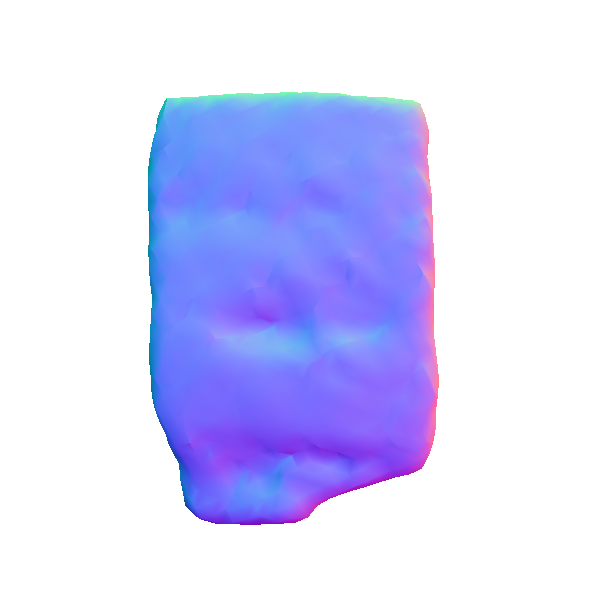} &
        \includegraphics[clip,width=0.07\linewidth]{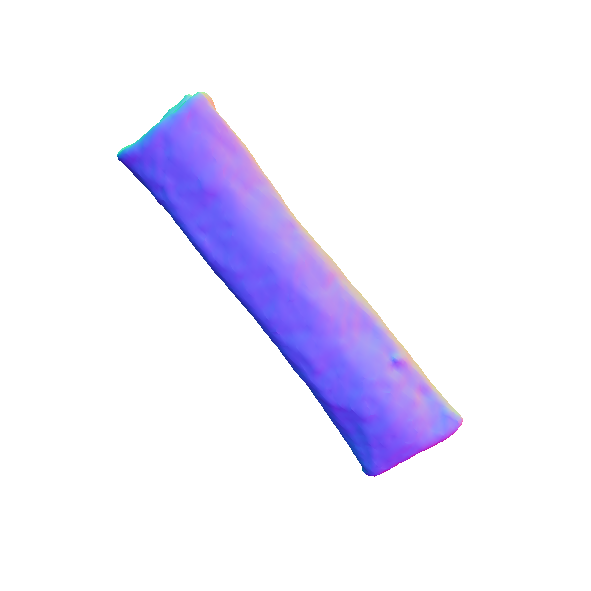} &
        \includegraphics[clip,width=0.07\linewidth]{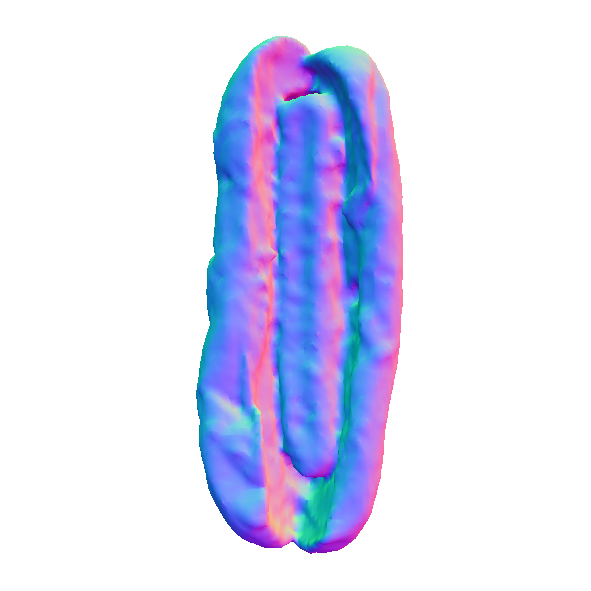}
        \\
        \hline
    \end{tabular}
    \caption{\change{Comparison between ours and VolETA reconstructions, and the ground-truth using the MTF dataset \cite{he2024metafood}. We use the labels in Table.\ref{table:quntitative_results} as user text prompts for our framework.}}
    \label{fig:qualitative_comparasions}
\end{figure*}

We extensively validated our approach on the MTF dataset and compared our results with the groundtruth meshes using MAPE and Chamfer distance metrics. More briefly, we leverage our approach for each food scene separately. A  food volume estimation is applied. PixSfM \cite{lindenberger2021pixel} estimates the poses and point cloud. 
Using the camera poses and the image, we run the text-guided segmentation by DEVA \cite{cheng2023tracking} to generate food-masked and reference object images. Afterward, we generate the meshes using NeRF and then apply the scaling factors to the given Food and Reference meshes. After generating the scaled meshes, we calculate the volumes and Chamfer distance with and without transformation metrics. We registered our meshes and ground truth meshes to obtain the transformation metrics using ICP \cite{rusinkiewicz2001efficient}.
Table \ref{table:quntitative_results} presents the quantitative volumes and Chamfer distance comparisons with and without the estimated transformation metrics from ICP using the MTF dataset. \change{We compare our approach with  VolETA \cite{almughrabi2024voleta,he2024metafood}, ININ. \cite{he2024metafood}, and FoodR. \cite{he2024metafood}, alongside ground truth volumes. The MAPE and Chamfer distance are also reported for each method.}
\begin{figure}[htb]
    \centering
    \setlength{\tabcolsep}{1pt}
    \begin{tabular}{ccc}
        Image & Reference & DEVA \\
        \includegraphics[width=0.16\textwidth]{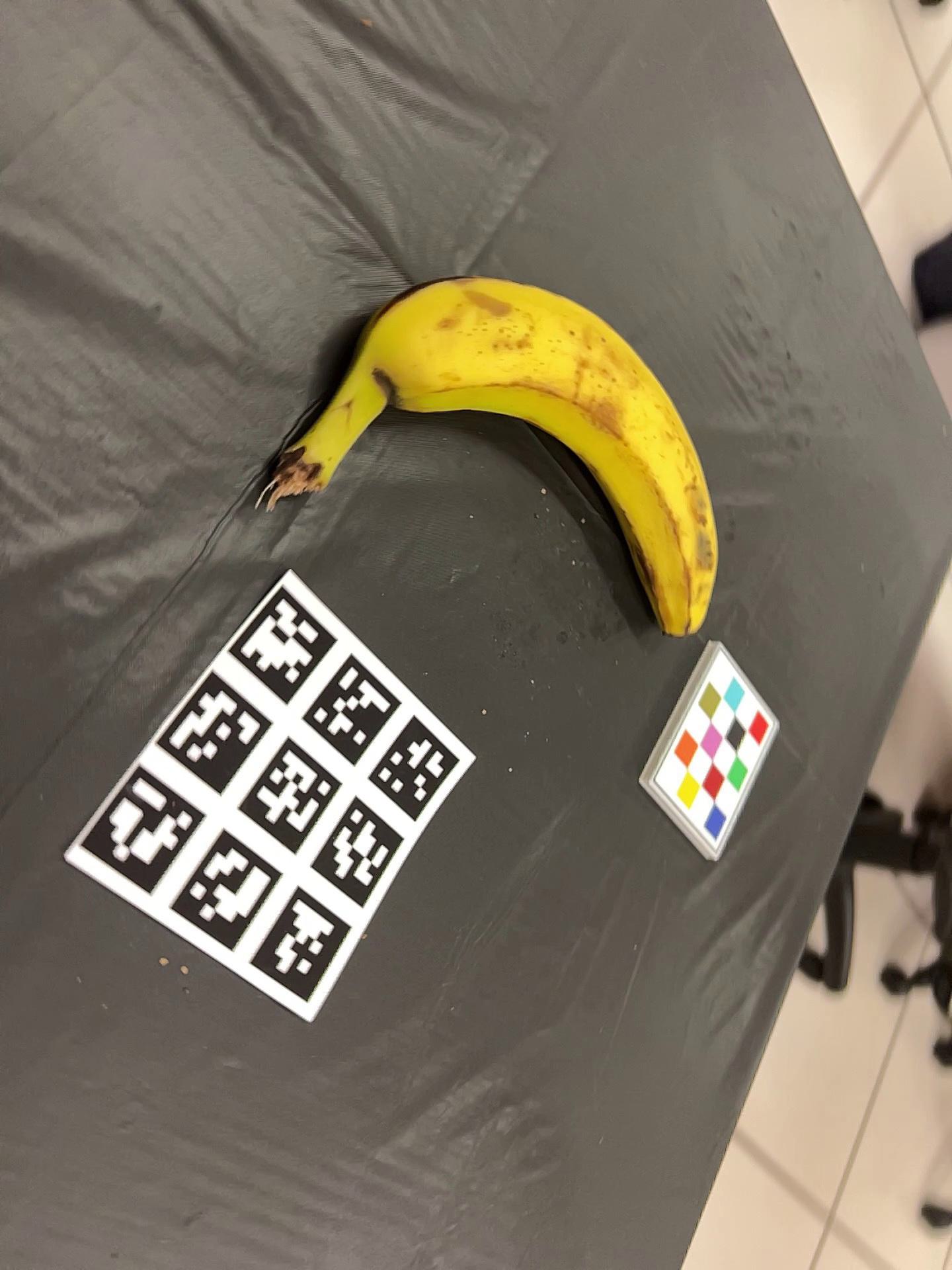}
        &
        \includegraphics[width=0.16\textwidth]{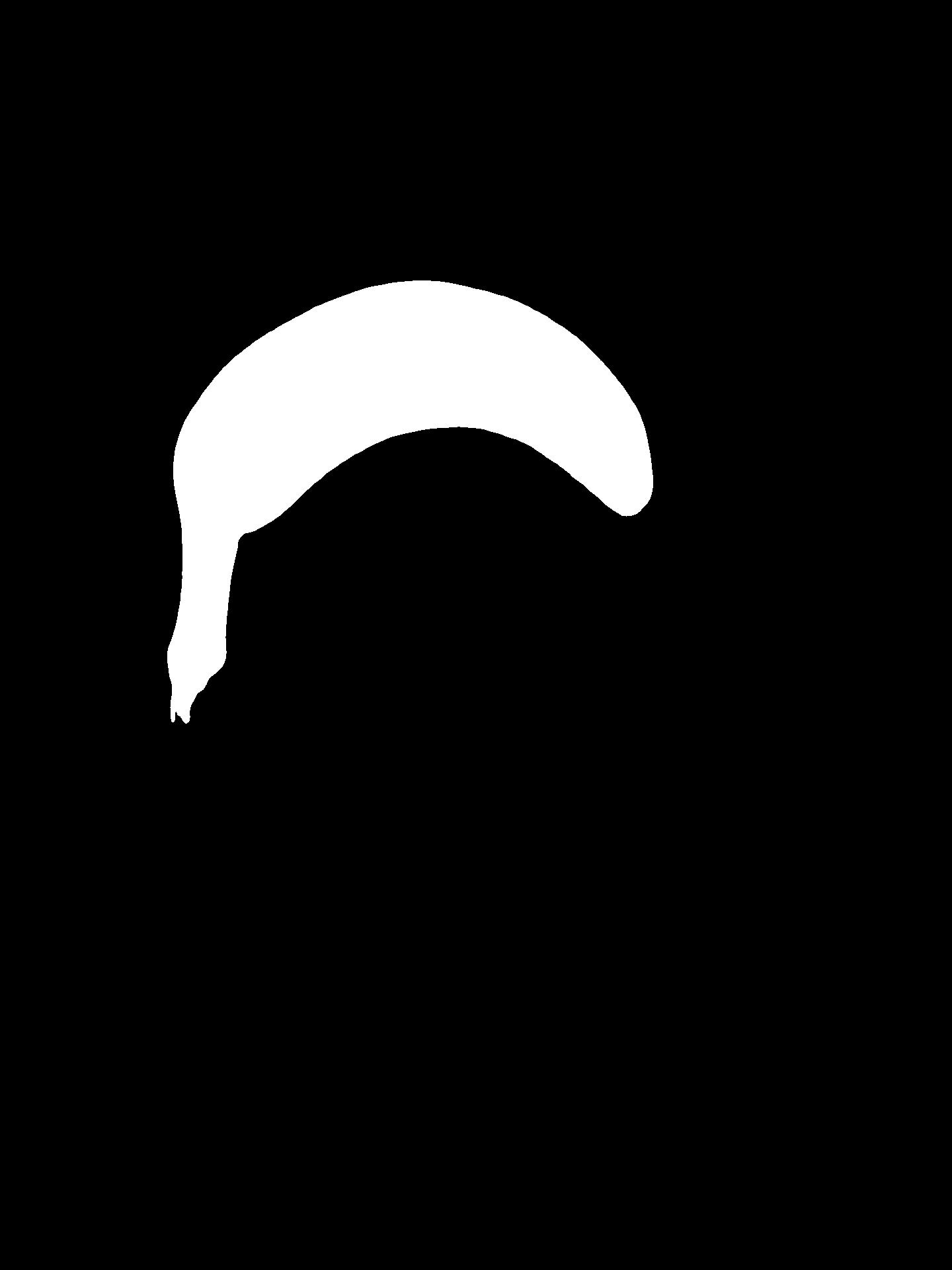}
        &
        \includegraphics[width=0.16\textwidth]{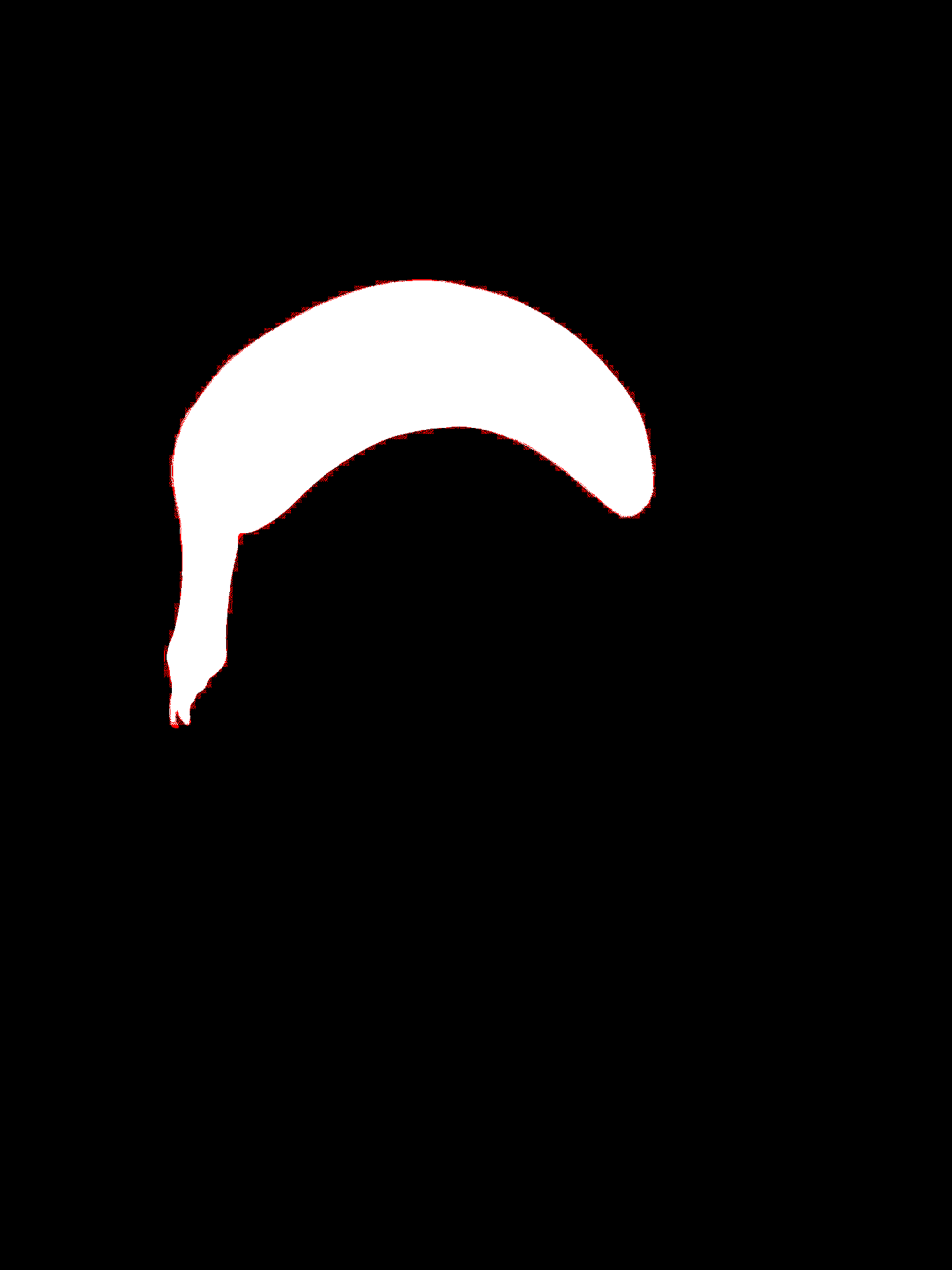}
        \\
        \includegraphics[width=0.16\textwidth]{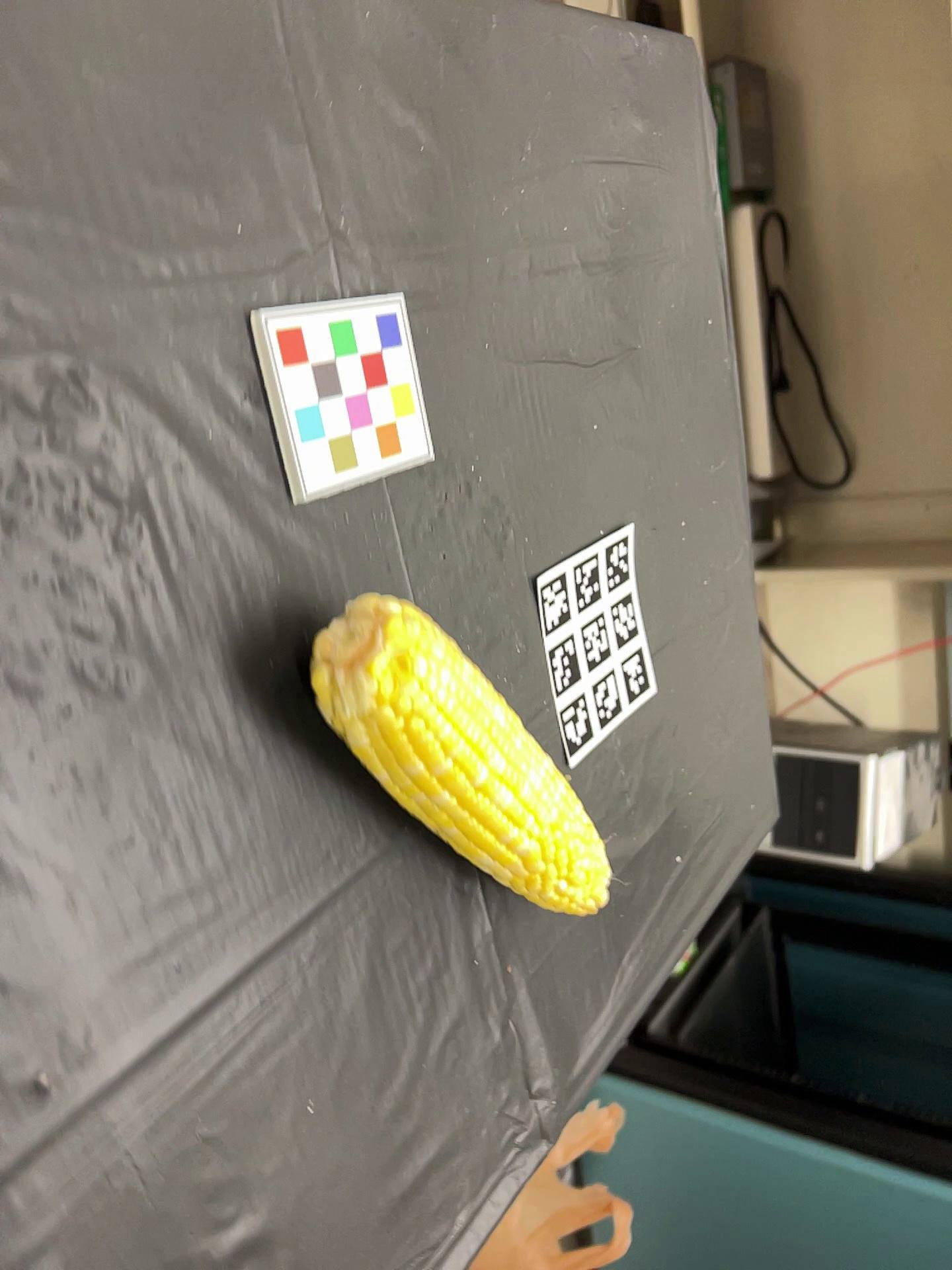}
        &
        \includegraphics[width=0.16\textwidth]{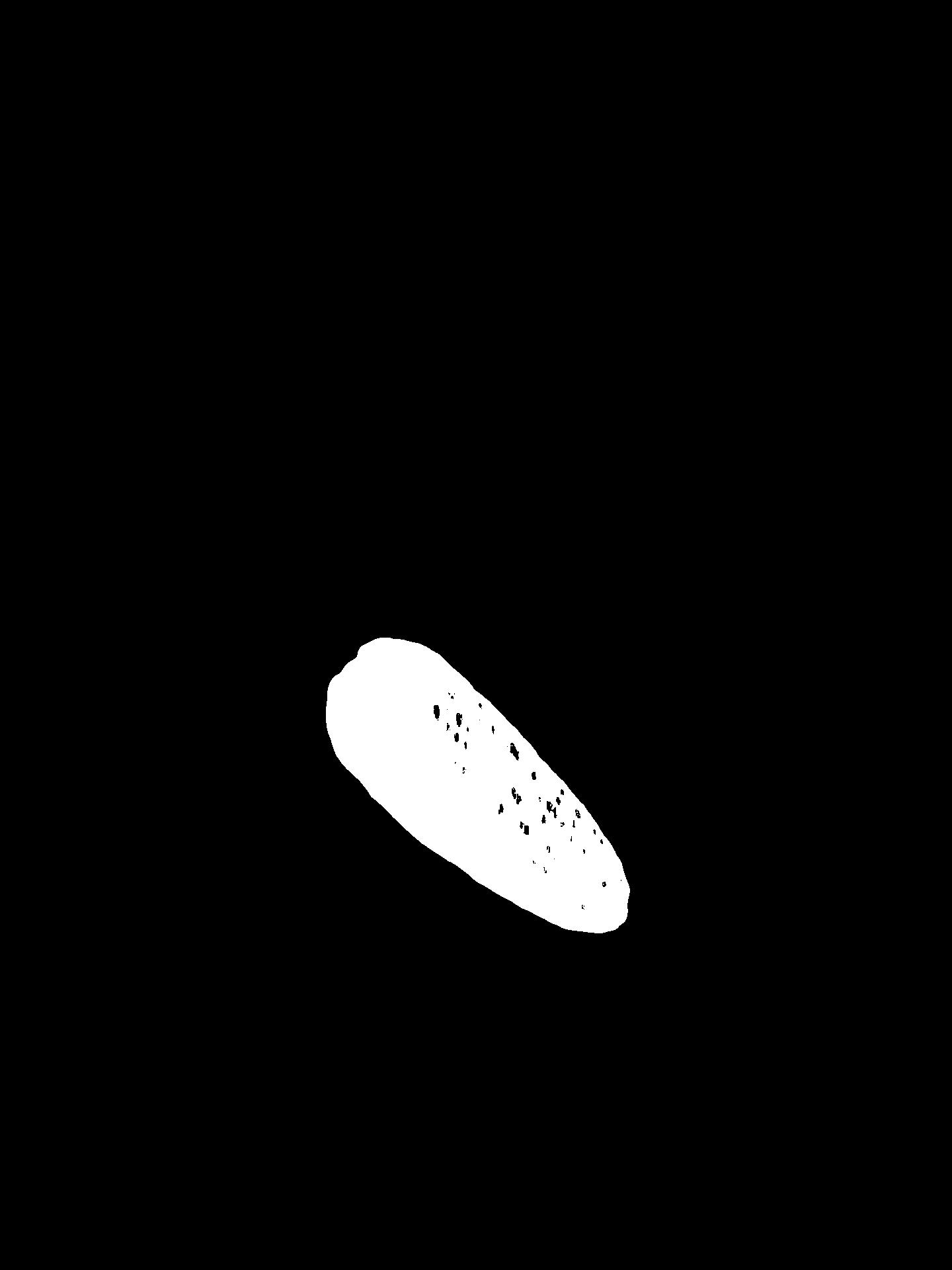}
        &
        \includegraphics[width=0.16\textwidth]{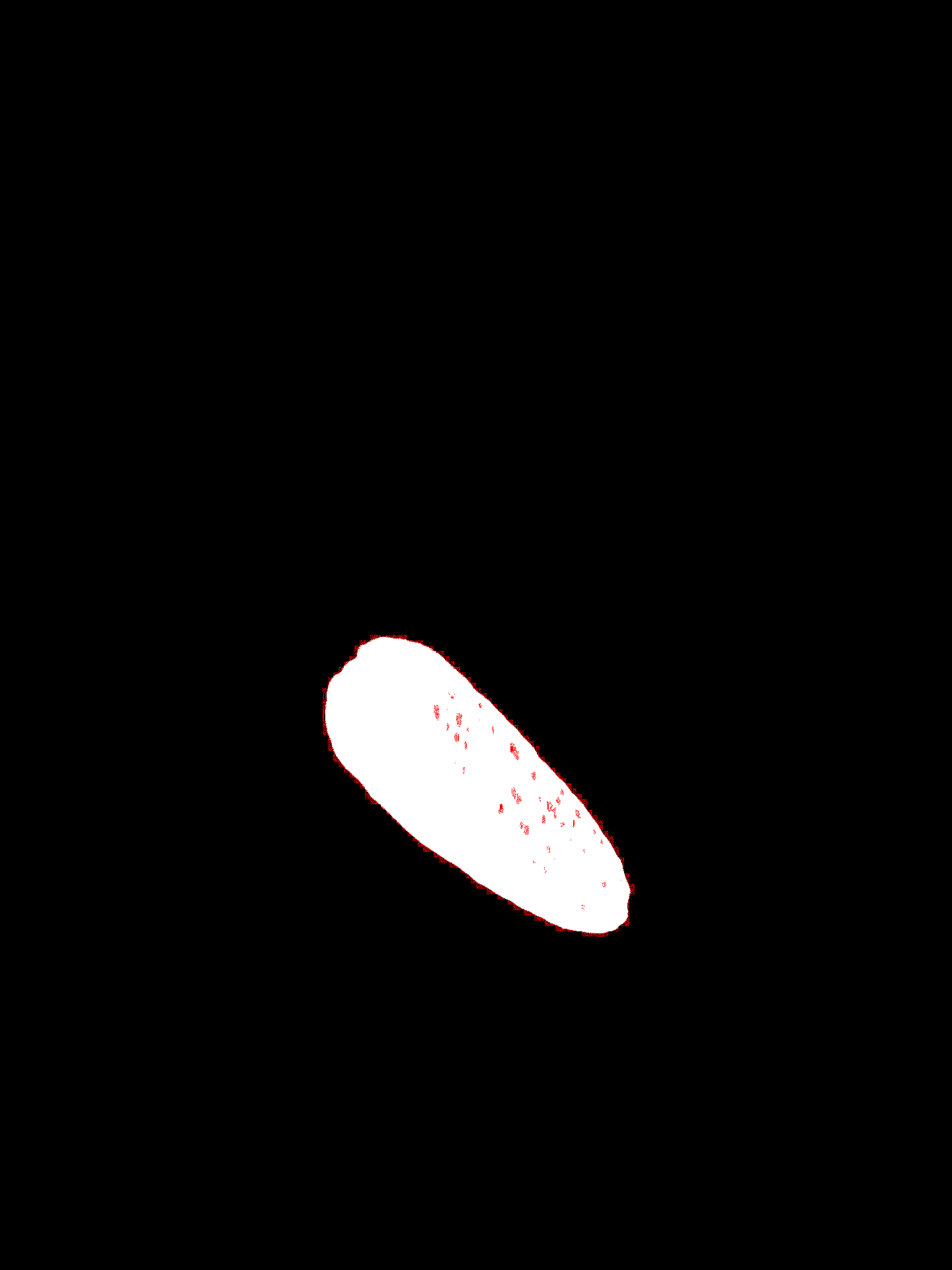}
        \\        \includegraphics[width=0.16\textwidth]{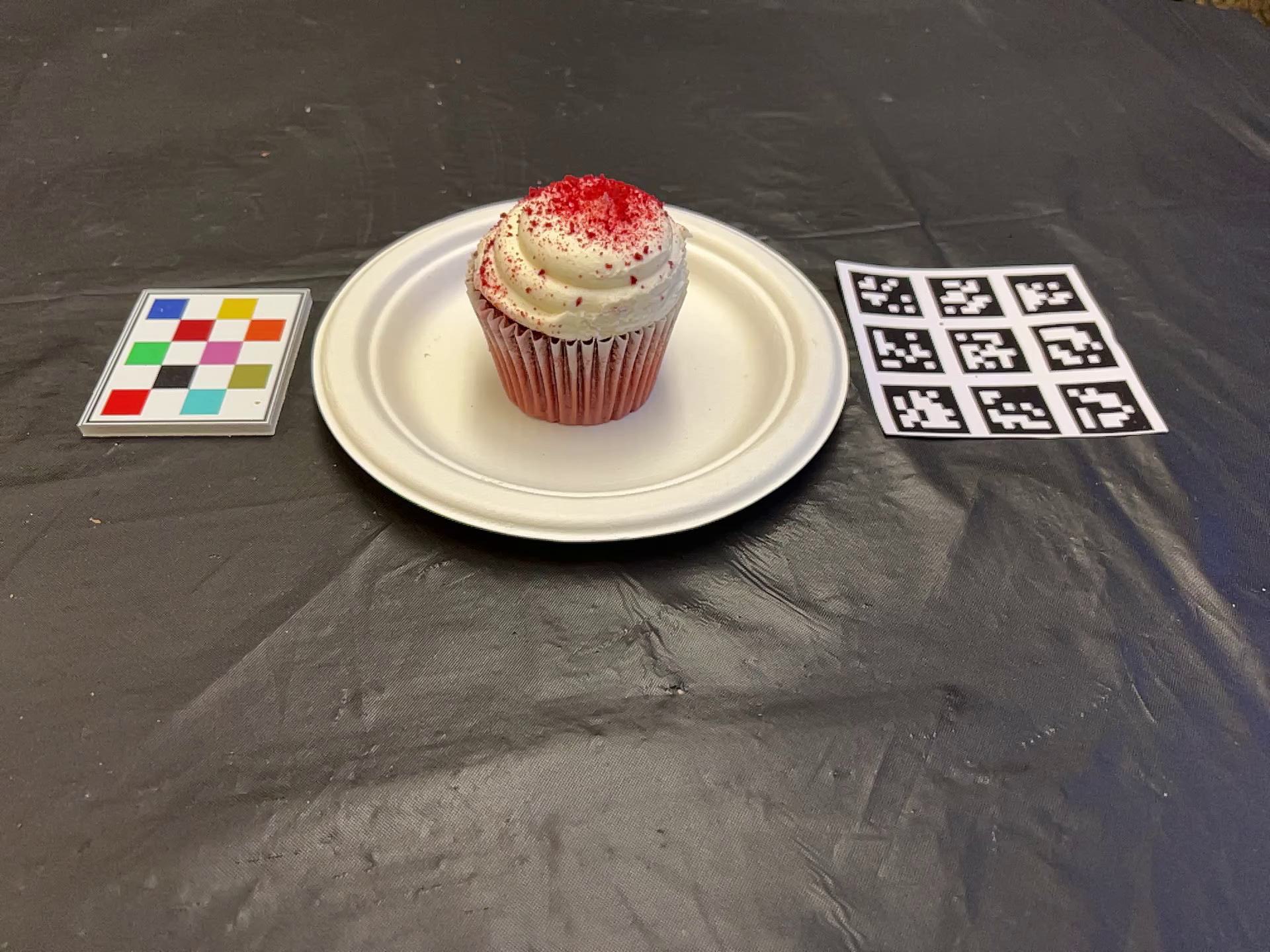}
        &
        \includegraphics[width=0.16\textwidth]{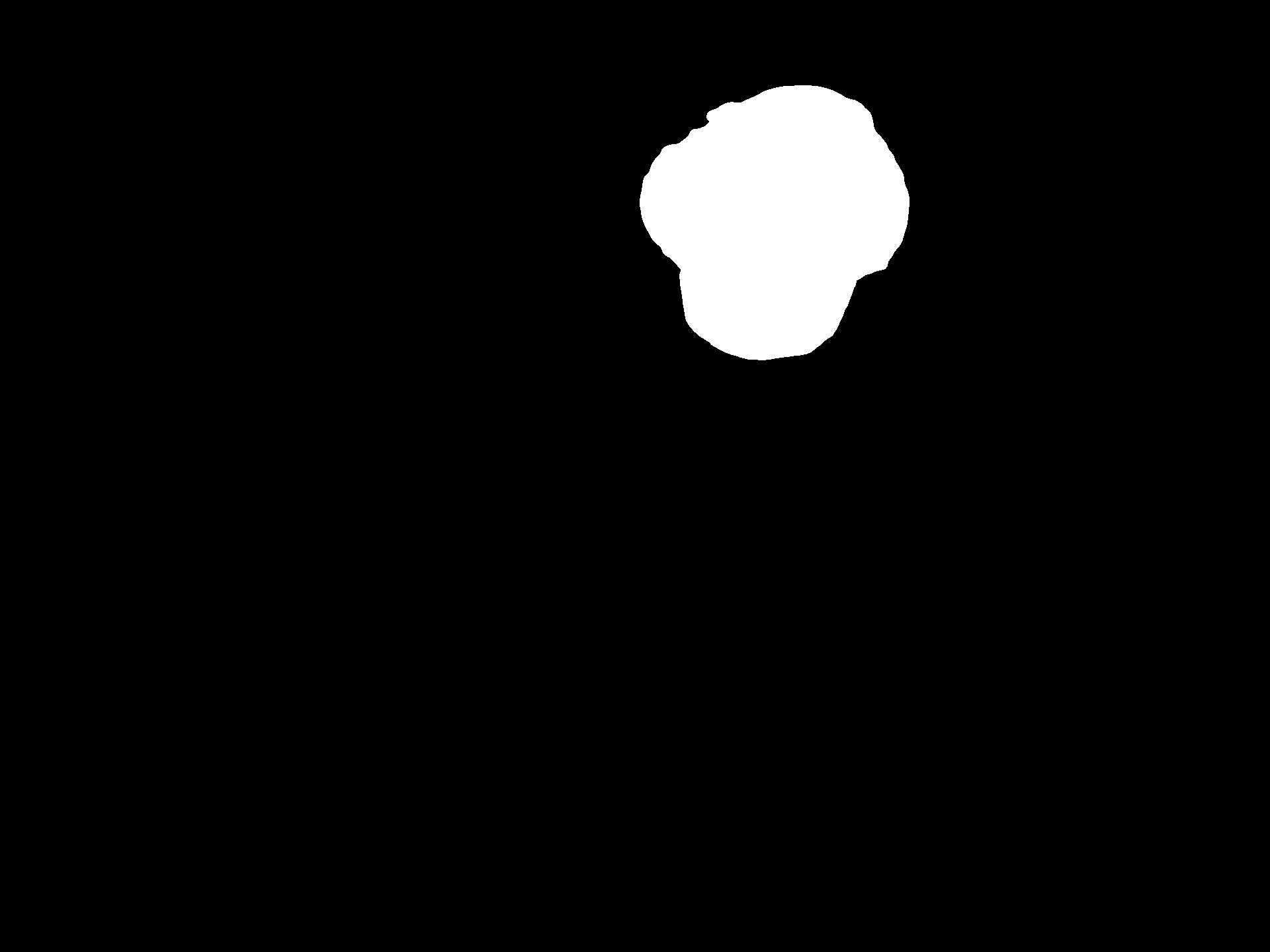}
        &
        \includegraphics[width=0.16\textwidth]{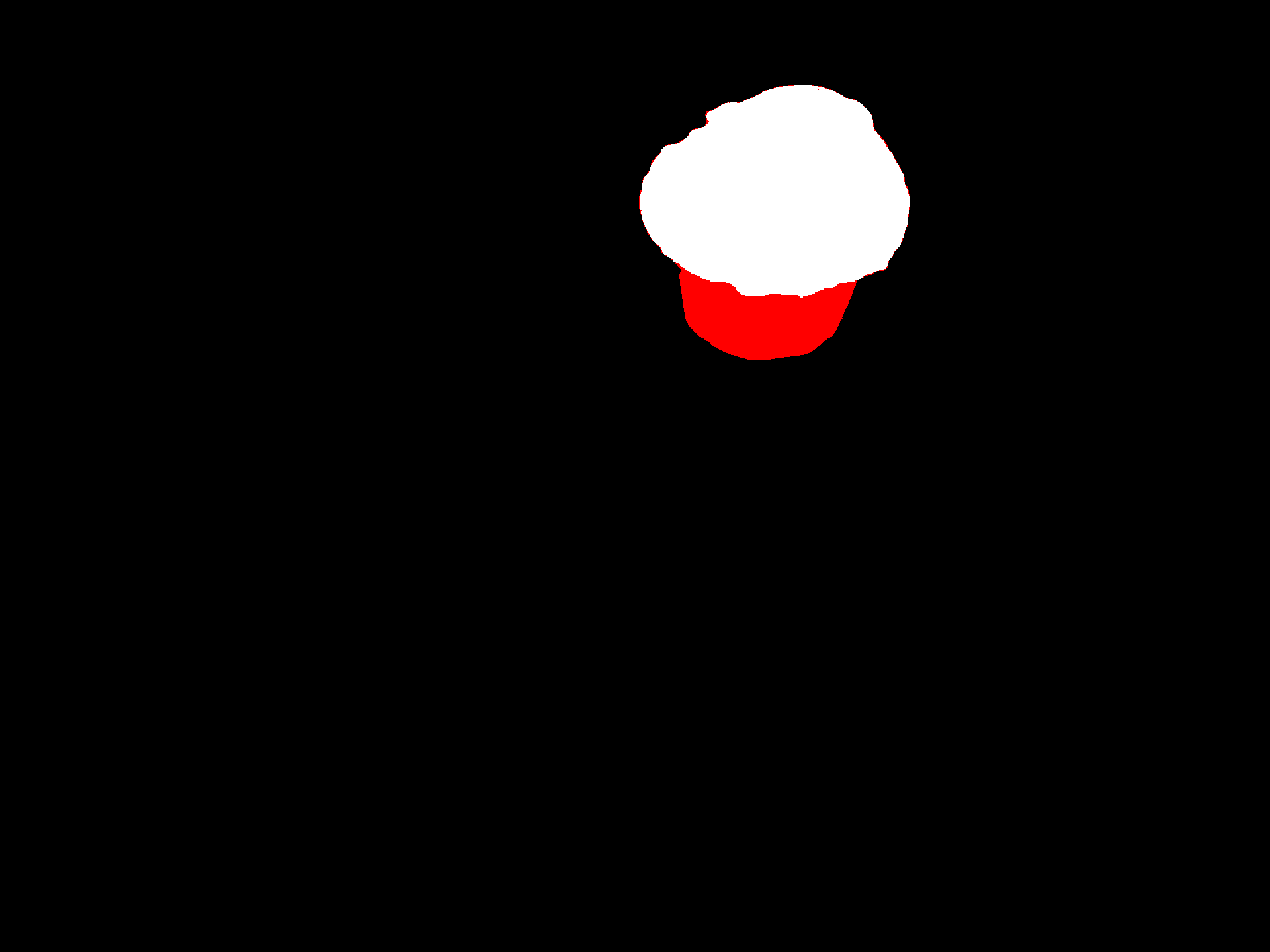}
    \end{tabular}
    \caption{\change{Comparison between original images, ground truth masks, and masks generated by DEVA. The dataset used is MetaFood3D. The red region highlights any artifacts or missing segments. We use the labels in Table \ref{table:quntitative_results} as user text prompts for our framework.}}
    \label{fig:Metafood3dMasksComparison}
\end{figure}
Additionally,  Fig.~\ref{fig:qualitative_results} and Fig.~\ref{fig:qualitative_comparasions} show the qualitative results on the 3D reconstruction from the MTF dataset. We examine the impact of scaling factors on 3D reconstruction methods, focusing on their influence on reconstruction quality and generalization from limited input data. Finally, Table~\ref{tab:scaling_factors_comparisions} shows additional experiments comparing scaling factors with baselines. \change{Our approach attains the lowest absolute error (0.0080) in scaling factor estimation, surpassing both VolETA and ININ (Reference). While the mean scaling factor estimated by our method (0.1061) exhibits close alignment with VolETA (0.1028), it demonstrates a substantially lower error in comparison to ININ, which reflects the highest deviation. Furthermore, our method achieves the lowest standard deviation (0.0123), indicative of greater consistency and robustness across various samples. The relative error metric further underscores the significant discrepancies in ININ's estimates, suggesting a reduced reliability for precise volume computations. These results substantiate that our approach yields the most accurate and stable scaling factor estimation, rendering it highly suitable for real-world food volume measurements.} \change{Fig.~\ref{fig:Metafood3dMasksComparison} illustrates the masks we have generated in comparison to the ground truth masks. Errors are emphasized in red; our masks exhibit a close resemblance to the ground truth, thereby indicating an exceptional segmentation of food object scenes.}

\paragraph{Discussion}
Accurate food volume estimation relies on segmentation, 3D reconstruction, and scaling factor computation. Text-guided segmentation plays a key role in reconstruction and optimizing the complexity of the pipeline, as shown in Table \ref{table:quntitative_results}, where Chamfer distance compares our text-guided segmentation with ground truth masks. ICP-based mesh registration influences these results due to alignment sensitivity. 3D reconstruction ensures structural consistency, with NeRF generating detailed meshes based on text-guided segmentation. Scaling factors convert unitless meshes into real-world dimensions, affecting final volume estimates (Table \ref{tab:scaling_factors_comparisions}).
Our approach enhances the controllability of NeRF-based reconstructions through text-guided segmentation, ensuring better adaptability to food volume estimation for the food scenes. 

\begin{table}[htb]
    \centering  
    \footnotesize
    \caption{Comparison of scaling factors and ININ. Approach against manually identified ground truth values. }
    \begin{tabular}{c|ccc|cc}
    \hline
    \multirow{2}{*}{ID} & \multicolumn{3}{c|}{Scaling Factors} & \multicolumn{2}{c}{Absolute Error↓} \\
    & \textbf{Ours} & VolETA(Ref.) & ININ. & \textbf{Ours} & ININ. \\
    \hline
1 & 0.0936 & 0.0896 & 0.0601 & 0.0013 & 0.0295\\
2 & 0.1093 & 0.1043 & 0.0818 & 0.0093 & 0.0225\\
3 & 0.0973 & 0.1043 & 0.0739 & 0.0089 & 0.0305\\
4 & 0.0977 & 0.0882 & 0.0836 & 0.0147 & 0.0046\\
5 & 0.1034 & 0.1034 & 0.0786 & 0.0024 & 0.0248\\
6 & 0.1267 & 0.1277 & 0.0884 & 0.0039 & 0.0393\\
7 & 0.1263 & 0.1043 & 0.1031 & 0.0194 & 0.0012\\
8 & 0.1039 & 0.1277 & 0.0685 & 0.0226 & 0.0592\\
9 & 0.0913 & 0.0876 & 0.0593 & 0.0003 & 0.0283\\
10 & 0.0907 & 0.0876 & 0.0582 & 0.0005 & 0.0294\\
11 & 0.1189 & 0.1043 & 0.0838 & 0.0143 & 0.0205\\
13 & 0.1097 & 0.1034 & 0.0697 & 0.005 & 0.0338\\
14 & 0.1099 & 0.1034 & 0.0738 & 0.0022 & 0.0297\\
\hline
Mean & 0.1061 & 0.0756 & 0.1028 & \textbf{0.0080} & 0.0272\\
Stdev. & \textbf{0.0123}	& 0.0129 & 0.0132 & \textbf{0.0075} & 0.0144 \\
\hline
Rel.$\dagger$ & \multicolumn{3}{c|}{} & - & 240\% \\ 
\hline
    \end{tabular}
    \label{tab:scaling_factors_comparisions}
    \vspace{-1em}
\end{table}


\paragraph{Limitations}
Our method is promising but has some limitations. 
The computational speed poses challenges since camera pose estimation and NeRF-based reconstruction \change{requires extensive computational process}.  
\change{A noteworthy limitation inherent in our methodology stems from segmentation errors, which result in incomplete mesh reconstructions that adversely impact volume estimation. As illustrated in Fig.~\ref{fig:Metafood3dMasksComparison}, the segmentation process inadvertently cropped a portion of the food mesh, leading to an underestimated volume in comparison to the established ground truth. This challenge is particularly pronounced for the food label "cake," where the segmentation process failed to preserve specific regions of the object, thereby generating a significant discrepancy in the predicted volume. This indicates that although our text-guided segmentation exhibits commendable performance in the majority of instances, it may result in incomplete reconstructions under certain circumstances.}

\section{Conclusions}
\label{sec:conclusion}
We presented VolTex, a framework for estimating food object volume that combines text-guided segmentation with NeRF-based 3D reconstruction. We evaluated our approach using the MTF dataset, which showed competitive results in food volume estimation. By leveraging text-guided segmentation, we enhanced NeRF \change{in estimating volumes for a specific object in food scenes}. Our future efforts will aim  \change{to enhance segmentation consistency} and maximize computational efficiency to speed up inference, making the framework more applicable in real-world scenarios.
\section*{Acknowldgements}
This work was partially funded by the EU project MUSAE (No. 01070421), 2021-SGR-01094 (AGAUR), Icrea Academia’2022 (Generalitat de Catalunya), Robo STEAM (2022-1-BG01-KA220-VET000089434, Erasmus+ EU), DeepSense (ACE053/22/000029, ACCIÓ), and Grants PID2022141566NB-I00 (IDEATE), PDC2022-133642-I00 (DeepFoodVol), and CNS2022-135480 (A-BMC) funded by MICIU/AEI/10.13039/501100 011033, by FEDER (UE), and by European Union NextGenerationEU/ PRTR. A. AlMughrabi acknowledges the support of FPI Becas, MICINN by PREP2022-000101, Spain. U. Haroon acknowledges the support of FI-SDUR Becas, MICINN, Spain.
{
    \small
    \bibliographystyle{ieeenat_fullname}
    \bibliography{main}
}


\end{document}